\documentclass[
nonbibnotes,
 amsmath,amssymb,
 aps,
twocolumn,
superscriptaddress,
]{revtex4-1}

\usepackage{graphicx}
\usepackage{dcolumn}
\usepackage[export]{adjustbox}
\begin{document}

\title{Holographic single particle imaging for weakly scattering, \\heterogeneous nanoscale objects}

\author{Abhishek Mall}
\affiliation{Max Planck Institute for the Structure and Dynamics of Matter, 22761 Hamburg, Germany}
\affiliation{Center for Free Electron Laser Science, 22761 Hamburg, Germany}

\author{Kartik Ayyer}
\email{kartik.ayyer@mpsd.mpg.de}
\affiliation{Max Planck Institute for the Structure and Dynamics of Matter, 22761 Hamburg, Germany}
\affiliation{Center for Free Electron Laser Science, 22761 Hamburg, Germany}
\affiliation{The Hamburg Center for Ultrafast Imaging, 22761 Hamburg, Germany}


\begin{abstract}

Single particle imaging (SPI) at X-ray free electron lasers (XFELs) is a technique to determine the 3D structure of nanoscale objects like biomolecules from a large number of diffraction patterns of copies of these objects in random orientations. Millions of low signal-to-noise diffraction patterns with unknown orientation are collected during an X-ray SPI experiment. The patterns are then analyzed and merged using a reconstruction algorithm to retrieve the full 3D-structure of particle. The resolution of reconstruction is limited by background noise, signal-to-noise ratio in diffraction patterns and total amount of data collected. We recently introduced a reference-enhanced holographic single particle imaging methodology [Optica {\bfseries7},593-601(2020)] to collect high enough signal-to-noise and background tolerant patterns and a reconstruction algorithm to recover missing parameters beyond orientation and then directly retrieve the full Fourier model of the sample of interest. Here we describe a phase retrieval algorithm  based on maximum likelihood estimation using  pattern search dubbed as MaxLP, with better scalability for fine sampling of latent parameters and much better performance in the low signal limit. Furthermore, we show that structural variations within the target particle are averaged in real space, significantly improving robustness to conformational heterogeneity in comparison to conventional SPI. With these computational improvements, we believe reference-enhanced SPI is capable of reaching sub-nm resolution biomolecule imaging.

\end{abstract}

\maketitle


\section{\label{sec:intro}Introduction}

Single particle imaging (SPI) experiments leverage X-ray free electron lasers (XFELs) to investigate the structure and dynamics of biological entities in their near-native state~\cite{Aquila:2015}. The ultrashort X-ray pulses enable the study of ultrafast structural dynamics in biological samples such as proteins and viruses at room temperature via diffraction before destruction~\cite{Chapman:2006}, and without the need to freeze samples. Many XFEL pulses with high flux and spatial coherence diffract from unknown target objects one at a time, in random orientations to collect millions of diffraction patterns. These are then computationally aligned and merged to reconstruct the 3D structure of the object~\cite{Sobolev:2020}. 

This analysis often broadly follows the following three steps: First, diffraction patterns containing signal from spurious contaminants, multiple particle agregates and other outliers are identified and removed using some machine learning framework~\cite{Yoon:2011,Ayyer:2021}. The rest of the patterns each represent a tomographic slice through the target object's 3D Fourier transform, albeit without the Fourier phases. The second step in the analysis pipeline is to align and average these patterns to obtain the 3D distribution of Fourier magnitudes~\cite{Loh:2009,Ayyer:2016}. Finally, iterative phase retrieval methods are used to recover the structure of the particle from these oversampled magnitudes, possibly with some background subtraction~\cite{Fienup:1978,Elser:2003,Lundholm:2018,Ayyer:2019}.

SPI has been used successfully in imaging samples in the 100-nm size range since they have high scattering cross section in comparison to the background, hence yielding high quality diffraction patterns~\cite{Ekeberg:2015,Rose:2018,Lundholm:2018}. However, for smaller particles, signal levels are much lower and extraneous background can severely hinder the alignment process. And while reconstruction algorithms are remarkably tolerant to low signal levels~\cite{Loh:2009,Philipp:2012,Ayyer:2019}, background often poses a fundamental limit on the achievable resolution since the diffraction signal from a compact object falls off very quickly with increasing momentum transfer but the background usually does not~\cite{Poudyal:2020}.

Various sample delivery methods have been used to deliver samples to the X-ray beam focus each balancing the requirements of maximizing efficiency (hit rate) with minimizing background. Aerosol sample delivery has low background signal but has a relatively low particle density, leading to low hit fractions~\cite{Munke:2016,Bielecki:2019}. Liquid-jets and solid substrates can be used as carrier media to increase the hit ratio~\cite{Chapman:2011,Sierra:2012,Hunter:2014,Nam:2016,Seuring:2018} but each hit now also has substantial signal from medium that obscures signal from the sample and limits orientation determination. With the advent of high repetition rate XFELs~\cite{Decking:2020}, even with the low hit rates of aerosol methods, one can still collect millions of patterns in a day~\cite{Ayyer:2021}. Nevertheless, even here, the background from the carrier gas of the aerosol and from detector false positives still produces sufficient background to limit the resolution to around 2 nm.

Recently, a reference-enhanced SPI~\cite{Ayyer:2020} technique was introduced, based on the holographic principle. The approach suggests attaching a strongly scattering particle to the target object to form a composite object. The references considered were a spherical gold nanoparticle (AuNP) and  a 2D crystal lattice with unit cell size comparable to target object. The total scattered signal is increased for each shot when a reference is attached and the signal now becomes more tolerant to background. A reconstruction algorithm was also developed to analyze the diffraction patterns generated from such a system in order to recover unknown parameters beyond orientation resulting from shot-to-shot variations in the composition of the composite object.

This holographic SPI technique alleviates the problem of high sensitivity to extraneous background, but at the cost of additional computational complexity in recovering the structure from the data. The technique introduces additional degrees of freedom in the composite system, which are unknown parameters for each pattern in diffraction dataset. The algorithm now has to recover not only the unknown orientation of the target object but also these hidden (latent) parameters characterizing the properties of the reference and the relative displacement between the reference and the target object. For the composite system where the reference is a spherical AuNP, the reference can be specified by one parameter: its diameter $D$. 

The previous work also introduced a reference-EMC algorithm~\cite{Ayyer:2020} which is a modified version of the EMC algorithm~\cite{Loh:2009,Ayyer:2016} that recovers these additional latent parameters and directly reconstructs the target object's complex Fourier transform and not just the intensities. However, as we will discuss in Section~\ref{sec:algo}, this method has problems dealing with very weak patterns as well as scaling to fine parameter sampling required to reach a high resolution.
In this paper, we propose a phase retrieval algorithm for holographic-SPI based on maximum likelihood estimation via pattern search dubbed as MaxLP. 
MaxLP scales efficiently with sampling of latent parameters and total number of diffraction patterns in comparison with previously introduced \emph{divide} and \emph{concur} approach in the reference-EMC algorithm~\cite{Gravel:2008}. Additionally, we observe that it performs much better when the signal level is low.
The algorithm is described in Section~\ref{sec:algo} and its performance is shown with simulated data in Section~\ref{sec:signaldep}.

We also discuss the performance of the holographic-SPI method with the MaxLP algorithm in the case where the target object itself is heterogeneous. In conventional SPI, this heterogeneity would have to be classified shot-by-shot before averaging since averaging intensities from variable structures can lead to meaningless results. Here, we show that in our approach, even without classification, we are able to reconstruct the average structure, making the process of handling conformational variations much more tractable. Thus, we find that this combination of the MaxLP algorithm with the holographic SPI experimental setup shows substantial promise in pushing the SPI technique to sub-nm resolution.

\begin{figure}
\begin{tabular}{cc}
\includegraphics[scale=0.67]{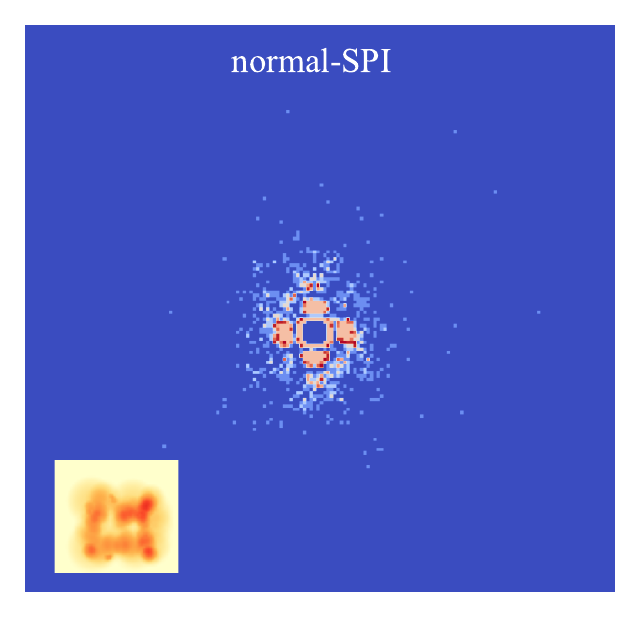}&\includegraphics[scale=0.67]{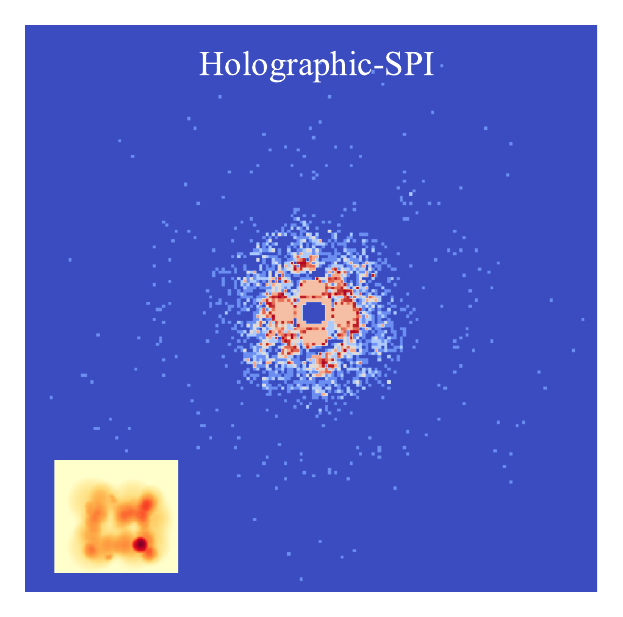} \\
(a) & (b)\\[6pt]
\end{tabular}
\caption{(a) Simulated diffraction pattern for a random object in a normal single-particle imaging (SPI) case with 4810 photons.  (b) Diffraction pattern from the same particle with a small AuNP attached (10,000 photons). If background is present, structural information can be discerned at higher scattering angles in the holographic case. Insets show the projected electron density of the objects.
}
\label{fig:frame}
\end{figure}

\section{\label{sec:algo}Reconstruction algorithm}
In a holographic SPI experiment, the sample of interest (target object) is in the vicinity of a strongly scattering reference, which in this work, we take to be a spherical gold nanoparticle (AuNP). Large number of diffraction patterns of these conjugates are collected in order to reconstruct the 3D structure of the target object. This reconstruction process consists broadly of two steps, the first being the determination of the unknown, or latent, parameters for each pattern, followed by a step to recover the structure given the data with the estimated parameters, which include the orientation of the object and the relative position and structure of the reference. Fortunately, the structure of the spherical AuNP can be described by just one number, namely its diameter $D$. For illustrative purposes, we limit ourselves to a two-dimensional object with only one in-plane rotational degree of freedom. Within this space, for a homogeneous reproducible object, we have four latent parameters to solve for: unknown orientation $\theta$ , diameter of AuNP $d$ and relative positions of AuNP and target object in $x$- and $y$- directions ($t_x$, $t_y$). In the general case, there are three orientational and translational parameters each, but the structure of the problem remains unchanged. The electron density of the composite object $\rho ({\bf r})$ is the sum of electron densities of the spherical AuNP $\rho _s({\bf r},D)$ and the unknown target object ${\rho _o}({\bf r})$
\begin{equation}
    \rho({\bf r}) = \rho_o({\bf r}) + \rho_s({\bf r- t}, D),
\end{equation}
where $\mathbf{t}$ is the relative shift of the centers of the two objects and $D$ represents the diameter of spherical AuNP. The total intensity distribution on the far-field detector in each frame is then
\begin{equation}
\label{eqn:intens}
    I({\bf q}, D, {\bf t}) = \left| F_o({\bf q}) + F_s({\bf q}, D)e^{2\pi i\mathbf{q.t}} \right|^2,
\end{equation}
where the $F$ terms represent the Fourier transform of electron densities, $F(\mathbf{q}) = \mathcal{F}[\rho(\mathbf{r})](\mathbf{q})$, and frame-by-frame shifts of the sphere transform to a phase ramp.

The simulated diffraction dataset contains of a large number of patterns defined by Eq.~\ref{eqn:intens}, each with random $D$ and $\mathbf{t}$ parameters and rotated in-plane by a uniform random angle $\theta$. These holographic intensities are then Poisson sampled with a given mean intensity level to generate the simulated photon counts per pixel. In Fig.~\ref{fig:frame}, we see the effect of the spherical AuNP attached as a reference to the target object on a simulated pattern for a given set of latent parameters with the same effective incident fluence. For the chosen AuNP size, the conjugates have on average twice as many total scattered photons with much of the excess in the higher order rings which improve the SNR in the presence of background at these resolutions. Larger AuNPs will scatter more strongly, but with lower contrast along the diffraction rings. The radii and intensity modulations in these rings are relevant to solve for diameter of the AuNP and its relative position. In order to retrieve the target object's structure which is assumed to be common to all the intensities, the first step is to solve for the set of latent parameters for each diffraction pattern and retrieve $F_o(\textbf{q})$. 

A modified version of EMC algorithm (Ref-EMC) was developed for holographic SPI~\cite{Ayyer:2020}. The conventional EMC algorithm involves three steps in each iteration: \textit{Expand}, \textit{Maximize} and \textit{Compress} which iteratively update the 3D intensity model to one which has a higher-likelihood of generating the observed diffraction patterns. When the reference is attached, the optimal final model is not the intensity distribution of the composite assembly, but rather the complex Fourier transform of just the target object, $F_o({\bf q})$. In the E-step of Ref-EMC, the latent parameter space is grid-sampled and predicted intensities are generated for each sample using the current estimate of $F_o({\bf q})$. In the M-step, the noisy diffraction patterns are compared with these predicted intensities and a probability distribution over parameters for each pattern is calculated. This distribution is then used to update the predicted intensities for each sampled parameter vector. The final C-step needs to recover the optimal $F_o(\mathbf{q})$ consistent with this stack of these intermediate predicted intensities, one for each sampled parameter set. This separate reconstruction problem is somewhat reminiscent of ptychography where multiple intensities are generated from a common object by varying translations~\cite{Pfeiffer:2018}. In this case however, the intensities are generated by coherent addition with a reference (Eq.~\ref{eqn:intens}) rather than by multiplication with a probe function. In Ref.~\onlinecite{Ayyer:2020}, a \textit{divide} and \textit{concur} iterative phase retrieval approach was implemented to solve this problem. However, the requirement to generate the intermediate intensities limits scalability of this algorithm. For a finer sampling of diameters and relative shifts, which is necessary for a high resolution, one ends up with many realizations of intermediate intensities which quickly becomes computationally expensive. Additionally, this method is composed of projection operations minimizing a Euclidean error metric. This implicitly assumes a Gaussian error distribution which becomes increasingly incorrect at lower signal levels~\cite{Thibault:2012}.

\subsection{Maximum-Likelihood  Phase Retrieval}
Taking the above mentioned considerations into account, we implement a maximum likelihood estimation strategy using pattern search technique dubbed as Maximum Likelihood Phaser (MaxLP) to retrieve the full $F_o({\bf q})$ of an unknown target object. The E- and M-steps of the EMC iteration are the same as before, with the only difference that we only use the most likely set of parameters rather than the whole probability distribution. The C-step is changed from the phase-retrieval-like approach using intermediate intensities to a direct search for the most likely complex Fourier amplitudes of the target object given the diffraction data and the current estimate of the latent parameters. The optimization is performed using a pattern search procedure implemented for each model pixel independently. The approach was tested for different intensity signal levels in diffraction datasets with low signal having as few as 2000 photons per pattern. We find that the algorithm is more noise-tolerant with better fidelity at lower signal levels where the previously introduced \emph{divide} and \emph{concur} approach yields poor results.

In addition, the MaxLP approach scales more favorably as the sampling of latent parameters is made finer. The computational complexity of the C-step is now independent of the sampling, while in the previous approach, finer sampling would result in a large number of intermediate intensities from which to perform the ``phase retrieval''. Since the final resolution is ultimately determined by the accuracy of the estimated latent parameters, this algorithm can be efficiently scaled and the parameters to be refined by local searches. 

\subsection{Single-pixel Behavior}

\begin{figure}
\begin{tabular}{cc}
\includegraphics[scale=0.6]{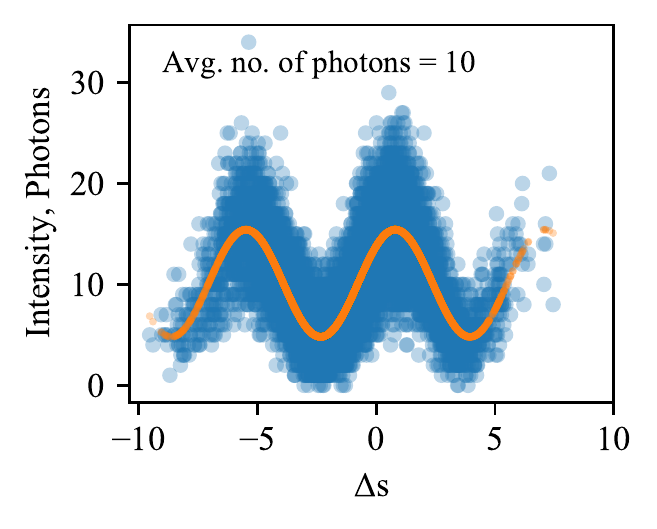}&\includegraphics[scale=0.6]{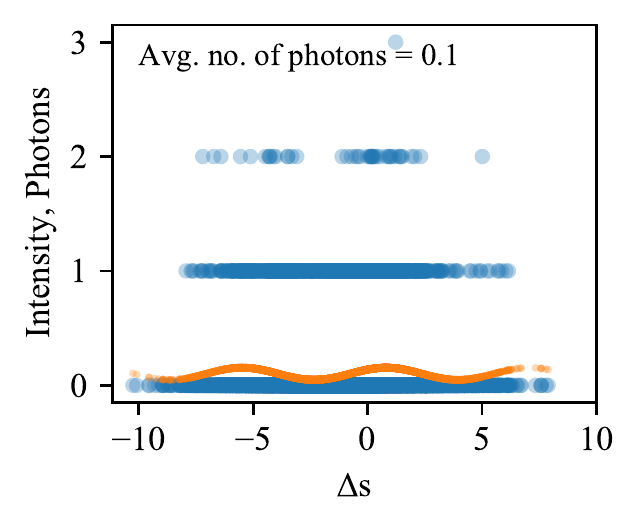}\\
(a) & (b)\\[6pt]
\includegraphics[scale=0.6]{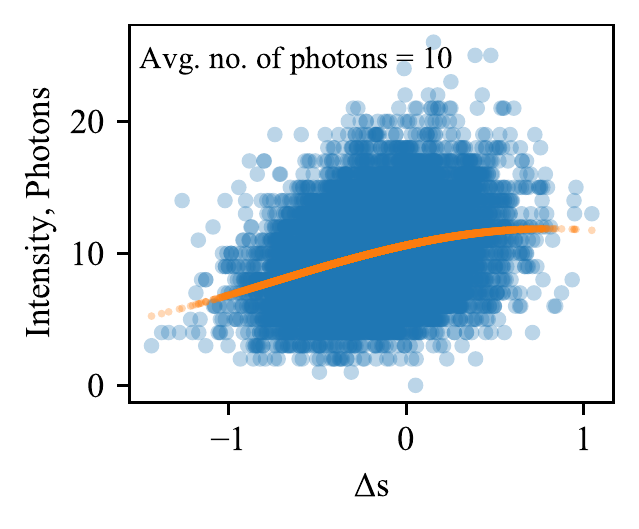}&\includegraphics[scale=0.6]{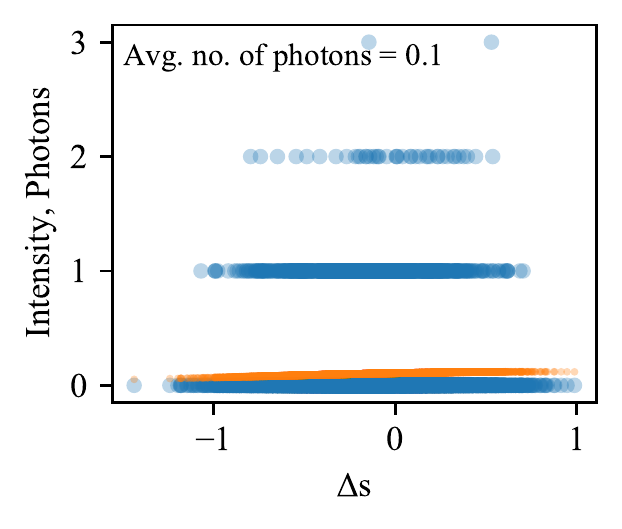}\\
(c) & (d)\\[6pt]
\includegraphics[scale=0.6]{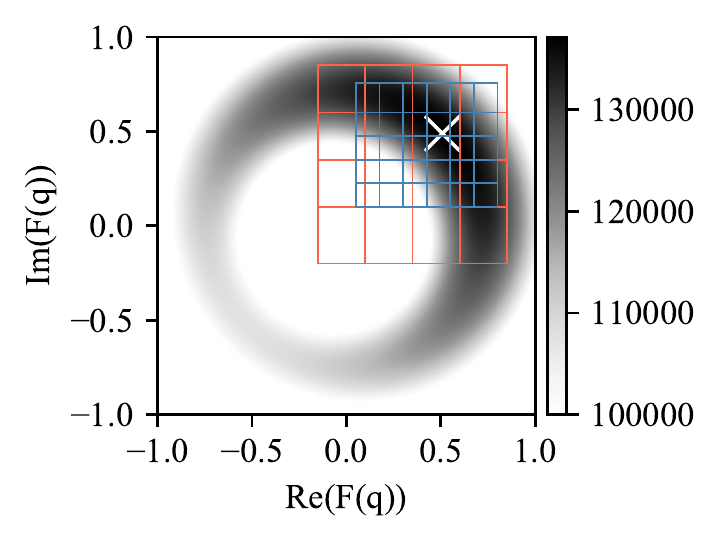}&\includegraphics[scale=0.6]{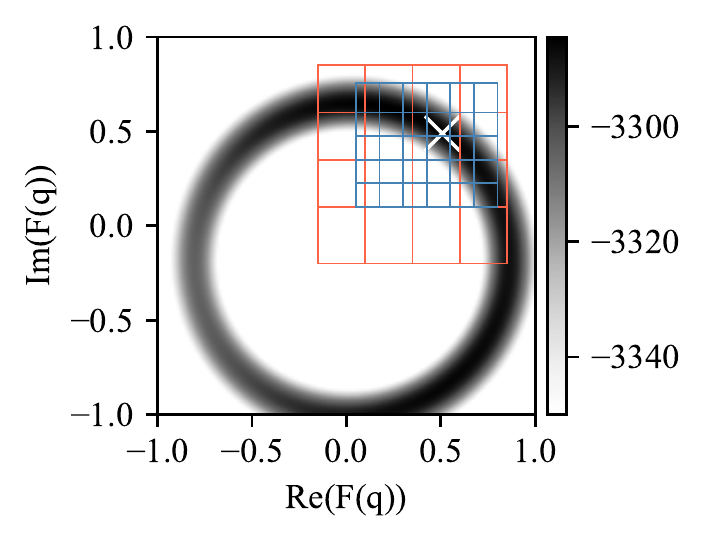}\\
(e) & (f)\\[6pt]
\end{tabular}
\caption{Intensity behavior at a single detector pixel for a holographic-SPI diffraction data. (a)-(d) Average number of photons \textit{vs} relative phase shifts of gold sphere with respect to the target object. The ideal intensity on the same pixel is represented by the orange line. (e) \& (f) Log-likelihood landscapes obtained from diffraction data for case (a) \& (d), respectively. The `$\times$' denotes the true value of $F(\mathbf{q})$ maximizing the log-likelihood. The grid lines ($1^\mathrm{st}$ - \emph{orange} and $2^\mathrm{nd}$ - \emph{blue}) depict the size of pattern search grid in consecutive iterations.}
\label{fig:pix}
\end{figure}

The optimal $F_o({\bf q})$ is determined at every model pixel independently. To illustrate the working of the reconstruction algorithm, we now discuss the behavior for a single pixel detector at a given $\mathbf{q}$ assuming the orientations are known. In holographic- or reference-enhanced-SPI, the measured intensity at a given detector pixel is described by Eqn~\ref{eqn:intens}. The estimated orientation of the object then relates the detector pixel to some pixel in the Fourier representation of the target object's electron density. 
A single model pixel intensity at a fixed \textbf{q} can be written as
\begin{equation}
\label{eqn:pix}
    I_\mathrm{pix} = \left|F_{o,\mathrm{pix}} + F_{s,\mathrm{pix}}(D) e^{2\pi i \mathbf{q}_\mathrm{pix}.\mathbf{t}}\right|^2,
\end{equation}
Here $F_o$ is the complex number which we need to solve from the diffraction dataset, $I_\mathrm{pix}$, with varying shifts, $\mathbf{t}$ and AuNP diameters, $D$, at the given model pixel from multiple realizations in different patterns.

Figure~\ref{fig:pix} depicts the behavior of the true intensity and measured photon counts for different phase shifts ($\Delta s =\mathbf{q}.\mathbf{t}$) between the AuNP and the target object. The  photon distribution at a given phase shift value is Poisson distributed with the mean being the true intensity shown in orange. When the average photon count at the pixel is high enough, the intensity follows a sinusoidal form with respect to the phase shifts, as seen in Fig.~\ref{fig:pix}(a). The relative phase, offset and amplitude of this sinusoidal curve tells us the phase and magnitude of the object's Fourier transform at that $\mathbf{q}$. As the range of phase shifts becomes small, one is effectively zooming into the sinusoidal curve for the same amount of average photons (see Fig.~\ref{fig:pix}(c)), and the fitting becomes more challenging. 
Similarly, when the average photon count on the detector falls, as seen in Fig.~\ref{fig:pix}(b), the photon distribution becomes too sparse to immediately see the true intensity and in the hardest case, for smaller phase shifts and low counts the true intensity plot becomes almost flat (Fig.~\ref{fig:pix}(d)).

The log-likelihood of such a diffraction dataset at any model pixel can be calculated by
\begin{equation}
\label{eqn:like}
     Q(F_{o,\mathrm{pix}}) = \sum_d K_{\mathrm{pix},d} log(I_{\mathrm{pix},d}) - I_{\mathrm{pix},d},
\end{equation}
where $I_{\mathrm{pix},d}$ is predicted intensity given by Eq.~\ref{eqn:pix} and $K_{\mathrm{pix},d}$ is the number of observed photons in pattern number $d$. The $K_{\mathrm{pix},d}!$ term is neglected since it does not depend upon the model $F_o$. Figure~\ref{fig:pix}(e-f) show this log-likelihood function distribution in the complex plane of $F_{o,\mathrm{pix}}$ for the diffraction pattern data given from Fig.~\ref{fig:pix}(a) and (d), respectively. The likelihood has a well behaved landscape with a sharp maxima with different values of real and imaginary part of $F_o$, when the average photon count is high enough with large shift range. When the average photon count is low with small shifts, the likelihood landscape becomes much flatter along the visible ridge.
 
Furthermore, this likelihood distribution is independent of the measured photon counts at the neighboring detector pixel.  As a result, one can implement an algorithm for the each pixel individually as a maximum-likelihood estimation problem and try to converge for an optimal $F_o$ for each pixel and trivially parallelize the algorithm over all model pixels. Of course, the Fourier transform is oversampled and thus, the model at neighboring pixels are strongly correlated. This information will be taken into account later in Section~\ref{sec:results}.

\subsection{Finding the most likely solution}
The likelihood function given in  Eq.~\ref{eqn:like}, can not be solved analytically. While one can perform this 2D optimization in the complex plane with many methods, we found that derivative based approaches were not robust, with maximization of likelihood failing for low photon count cases and small shifts. We found reasonable success with a pattern search approach~\cite{Torczon:1997} for two reasons: it could be efficiently parallelized on the graphical processing units (GPUs) and secondly it was quite robust in low signal and small shift cases.

Pattern search optimization is essentially a non-derivative technique that does not require a gradient calculation in its update of parameters. The search begins with a 2D grid of values of \textbf{F} (see Fig.~\ref{fig:pix}(e-f), orange grid lines). At each iteration, this grid moves to a set of values which best maximizes the likelihood function. If it finds a set of values of \textbf{F} which does not have better likelihood then it stays at current value and the grid shrinks and becomes denser with smaller steps between the grid values (blue grid lines). A search is run till a threshold error between the current and the previous estimate is reached.
This is performed for all the model pixels and then the final model is used to get updated latent parameters for the next EMC iteration.

\section{\label{sec:results}Results}

We tested the performance of the algorithm described above on 2D simulated data with a single angular degree of freedom. The target object was a randomly generated agglomeration of small spheres, approximating the electron density distribution of a biological sample. Each diffraction pattern was a Poisson-sampled distribution of scattered photons from a conjugate object consisting of the target attached to a spherical gold nanoparticle (AuNP) reference. The average diameter of the AuNP was roughly 1/5th the size of the target and the contrast was 11 times higher, to reflect the electron density ratio of gold to organic matter. The conjugate objects varied from pattern to pattern in multiple ways. The relative shifts in $x$- and $y$- directions between their centers were sampled from a normal distribution with a standard deviation of 1 pixel. The diameter of AuNP was also normally distributed with a mean and standard deviation of 7 and 0.5 pixels, respectively. The intensity distribution from this composite object was rotated in-plane by a uniform random angle before Poisson sampling. The diffraction patterns were collected on a circular detector with a diameter of 185 pixels and a central hole with a 4 pixel radius.

Multiple datasets were simulated with a varying signal levels with 10,000 patterns in each dataset. A common randomly generated object was used as the target in all of the simulations. 
The signal levels ranged from $2\times10^3$ to $10^5$ photons/frame at the low and high extremes.

Figure~\ref{fig:intens}(a) and (b) depicts the absolute magnitude of the reconstructed complex Fourier model for the highest and lowest signal levels. At the final reconstruction iteration, both Fourier models have a few poorly reconstructed pixels, primarily at low $q$. This is due to the missing data in the central hole as well as regions where the scattering from the reference is significantly weaker than from the target. The intensity at these pixels is only minimally affected by the frame-to-frame variation in shifts or diameters, resulting in the Fourier phase being poorly constrained. 
These pixels were filled in using the following approach. First, a support mask was calculated using data from moderate $q$ using the dark-field diffractive imaging approach~\cite{Martin:2012}. This mask was then used as a real-space constraint and the difference map iterative phase retrieval algorithm~\cite{Elser:2003} was used to fill in the missing region as well as obtain a real-space image.

\subsection{\label{sec:signaldep}Signal level dependence of reconstruction quality}
\begin{figure}
\begin{tabular}{cc}
\includegraphics[scale=0.65]{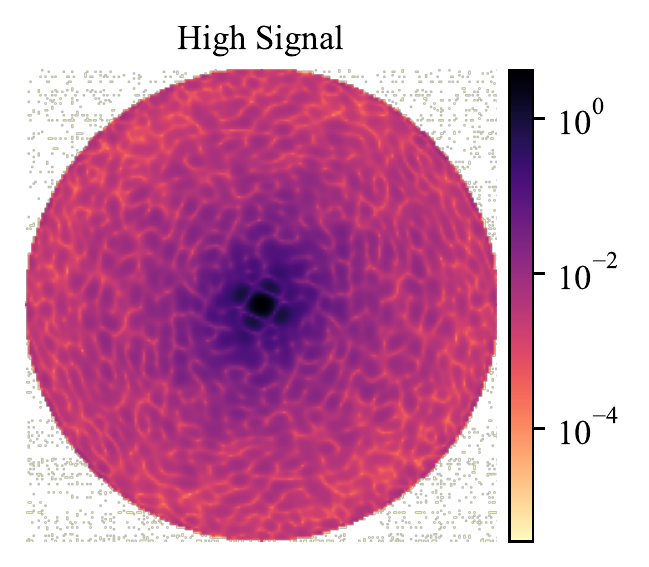}&\includegraphics[scale=0.65]{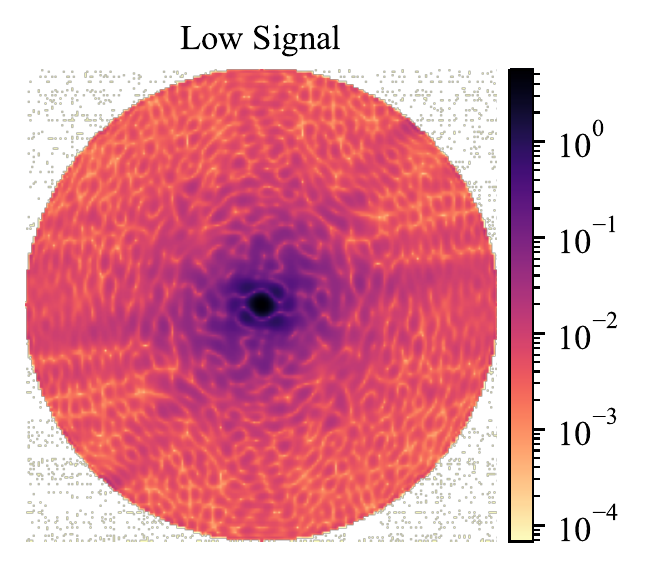}\\
(a) & (b)\\[6pt]
\end{tabular}
\begin{tabular}{c}
\includegraphics[scale=0.67]{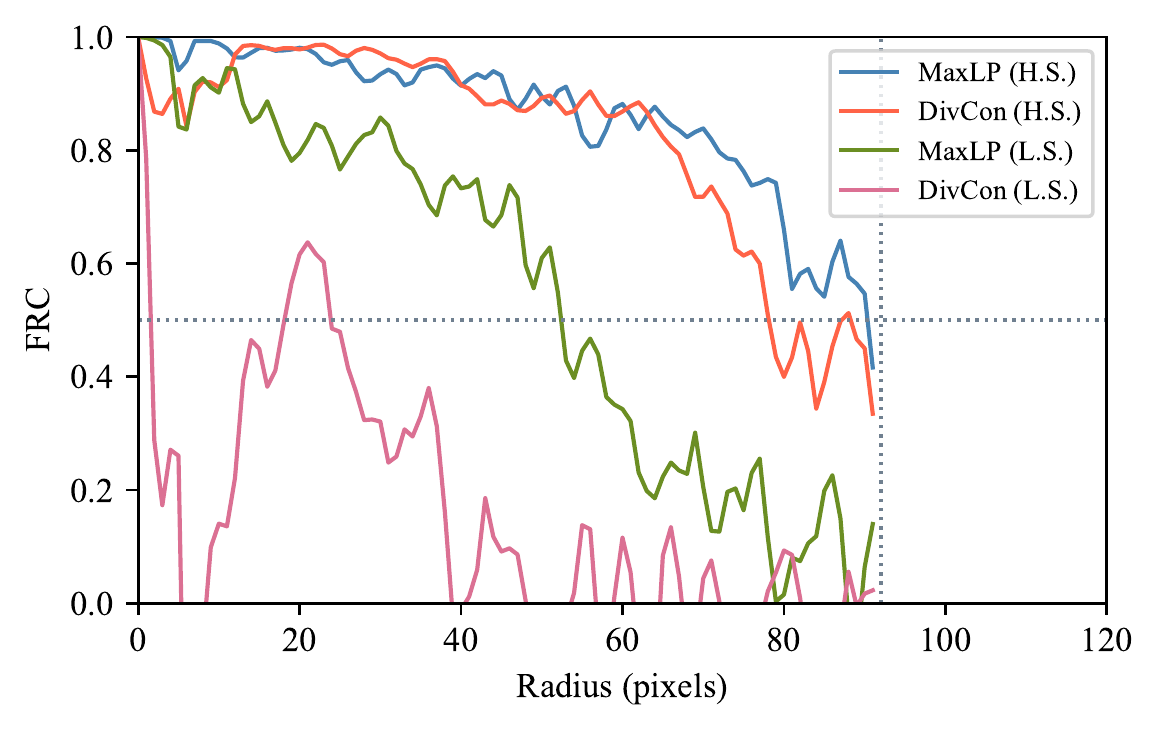}\\
(c)\\[6pt]
\end{tabular}
\caption{Simulation results for homogeneous target object attached to a spherical AuNP as reference. (a) Magnitude of reconstructed Fourier model for high signal diffraction data with MaxLP method ($10^5$ ph/frame). (b) Same for low signal diffraction data (2000 ph/frame). Both datasets had $10^4$ frames. (c) Comparison of Fourier ring correlation (FRC) between reconstructions and ground truth with different signal level diffraction data. Maximum-likelihood Phaser (MaxLP), Divide and Concur (DivCon), High Signal (H.S.) and Low Signal (L.S.)}
\label{fig:intens}
\end{figure}

The fidelity of the reconstructed model is calculated using the Fourier Ring Correlation (FRC) metric, shown in Figure~\ref{fig:intens}(c). We compare the performance of MaxLP method with the previously published method of \emph{divide} and \emph{concur} in [\onlinecite{Ayyer:2020}] at low and high signal levels. For high signals, both algorithms perform well, but the \emph{divide} and \emph{concur} approach fails for the low signal case, whereas the MaxLP reconstructs the Fourier model with FRC $>0.5$ up to $q = 55$ pixels. This can be attributed to the MaxLP method correctly accounting for the Poisson noise process rather than the Gaussian noise model implicit in the \emph{divide} and \emph{concur} method.


\begin{figure}
\begin{tabular}{cc}
\includegraphics[scale=0.6]{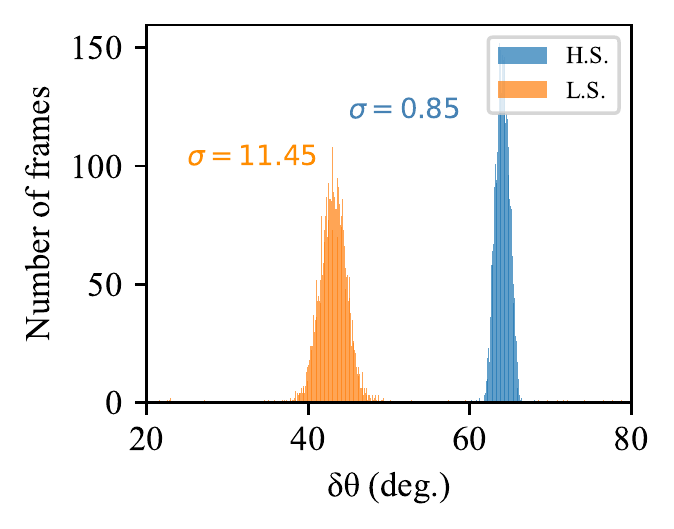}&\includegraphics[scale=0.6]{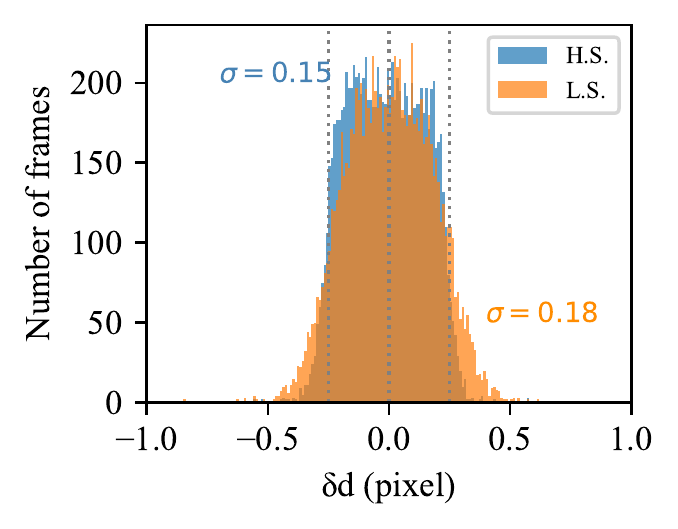}\\
(a) & (b)\\[6pt]
\includegraphics[scale=0.6]{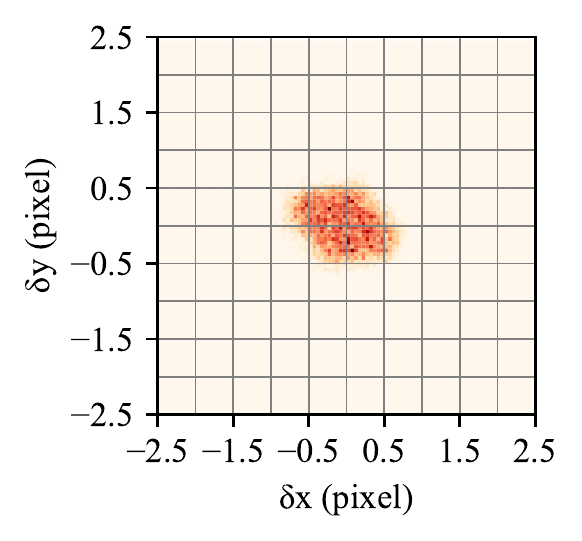}&\includegraphics[scale=0.6]{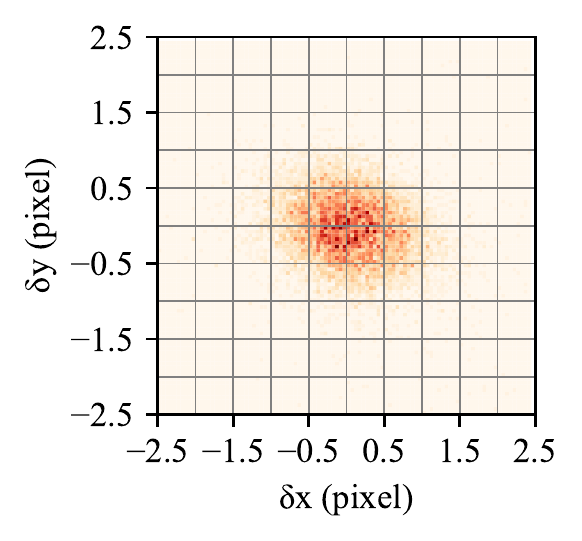} \\
(c) & (d)\\[6pt]
\end{tabular}
\caption{Error distribution for reconstructed latent parameters for homogeneous target object attached with a spherical AuNP as reference for low (L.S.) and high signal (H.S.) levels. (a) In-plane orientation, (b) diameter of gold sphere, (c) \& (d) relative  displacements between center of target object and reference in $x$- and $y$- direction. The vertical lines in (b) and the grid lines in (c) and (d) show half the sampling rate, which sets a lower bound on the width of the distributions.}
\label{fig:param}
\end{figure}

We then investigate the ability of the algorithm to correctly estimate the latent parameters given the current model. The sampling rate of all the latent parameters is kept identical across all datasets to facilitate comparison. Figure~\ref{fig:param} shows the error distribution for the diameter of AuNP , in-plane orientation and relative shifts in $x$- and $y$- directions, respectively. The sampling of the in-plane orientation is in range of 0 and $180^\circ$ with step size of $2^\circ$. In Fig.~\ref{fig:param}(a) one can see distribution of errors between the retrieved and true in-plane angles for high and low signal level, respectively. The centers are shifted since the reconstruction has an arbitrary overall orientation. As expected, the distribution is narrow for high signal, with $\sigma = 0.85^\circ$ which is below the sampling rate, and significantly broader for low signal with $\sigma = 11.4^\circ$.

Similarly, AuNP diameter and $x-$ $\&$  $y-$ shifts are sampled with a step size of 0.5 and 1 pixel, respectively. Figure~\ref{fig:param}(b) shows that the diameter of the AuNP for each frame is close to the true values with uncertainties for high and low signal being 0.15 and 0.18 pixels. Even for the lowest signal level, the diameter can be accurately estimated below half the sampling rate, which is consistent with the high precisions which can be obtained in small-angle X-ray scattering. Figure~\ref{fig:param}(c) $\&$ (d) show the error distribution of shifts in $x$- and $y$- direction for the two signal levels. The error is confined for many of the patterns within half the sampling bin size of 1 pixel. However the error distribution for the low signal case is relatively broad. This is strongly correlated to the error in the in-plane orientations. A large error in orientation leads to wrongly estimated shifts because the diffraction pattern is obtained by adding the AuNP to the target object and rotating the entire composite object. The predicted shift values are rotation-corrected using the predicted orientations and then compared with true values.

\begin{figure}
\begin{tabular}{cc}
\includegraphics[scale=0.6]{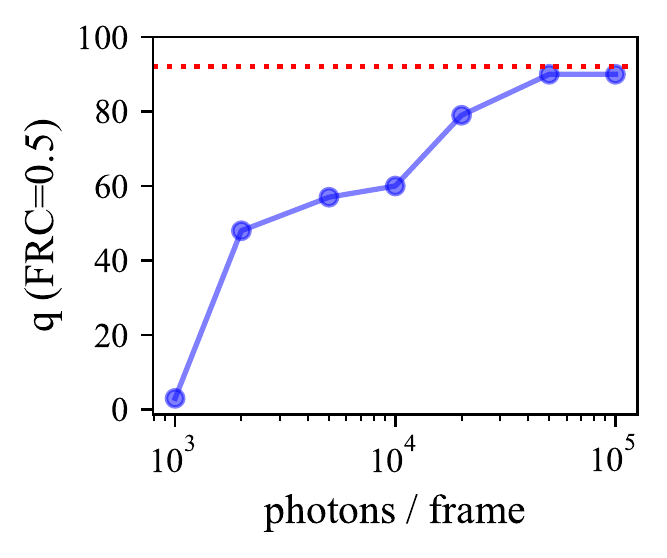}&\includegraphics[scale=0.6]{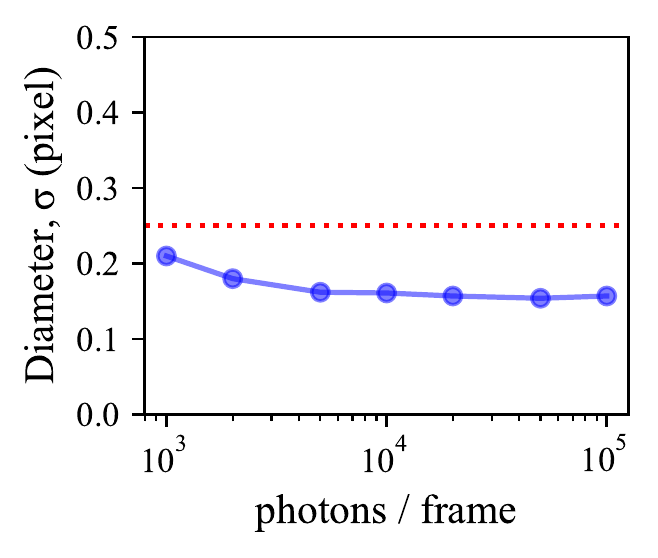}\\
(a) & (b)\\[6pt]
\includegraphics[scale=0.6]{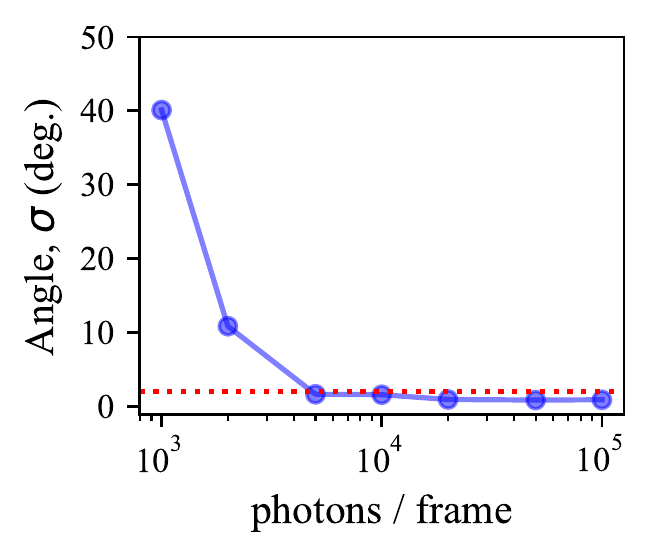}&\includegraphics[scale=0.6]{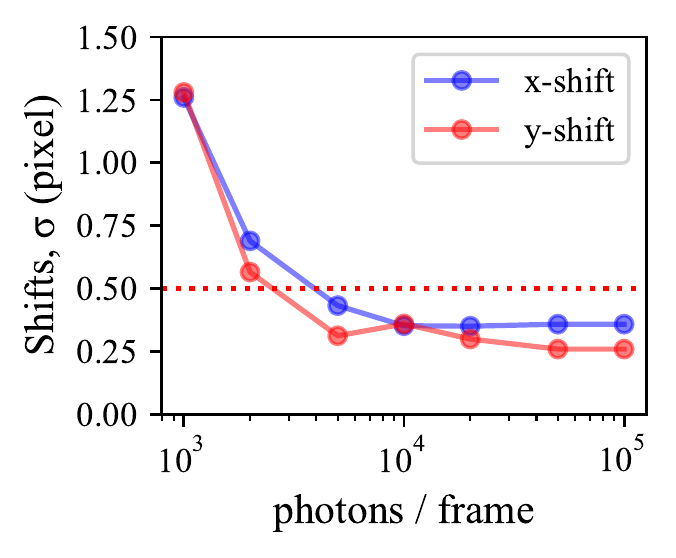} \\
(c) & (d)\\[6pt]
\end{tabular}
\caption{Evaluation metrics of Maximum-likelihood phaser (MaxLP) for different photon signal levels. (a) $q$ value when FRC = 0.5 \textit{vs} number of photons/frame. The horizontal dashed line  corresponds to a resolution of 1 real-space pixel. Standard deviation values for (b) diameter errors, (c) orientation errors, and (d) shift errors in $x$- and $y$- direction. The horizontal dashed lines in (b)-(d) indicate half the sampling rate. }
\label{fig:err}
\end{figure}

The behavior of the reconstruction metrics discussed above is evaluated for multiple intermediate signal levels in Fig.~\ref{fig:err}. Figures~\ref{fig:err}(b-d) show the dependence of the latent parameter errors on the signal level. Here we see that beyond 5000 photons/frame, the errors are all below the sampling rate indicated by the dotted red line. Figure~\ref{fig:err}(a) shows the dependence of the resolution on signal level, which depends both on accurate estimation of latent parameters as well as total signal from all the frames in aggregate. Thus, we expect the reconstruction FRC to improve significantly with more patterns beyond this threshold of 5000 photons/frame.

\subsection{Heterogeneity of target object}

\begin{figure}
\begin{tabular}{ccc}
\includegraphics[scale=0.67]{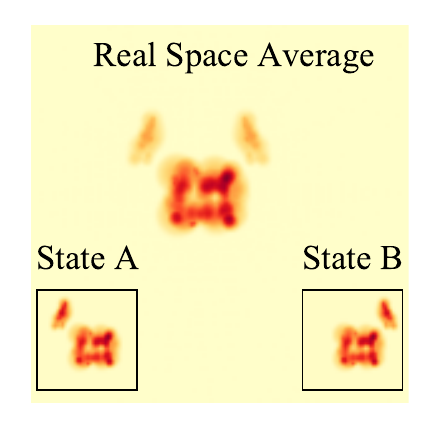}&\includegraphics[scale=0.67]{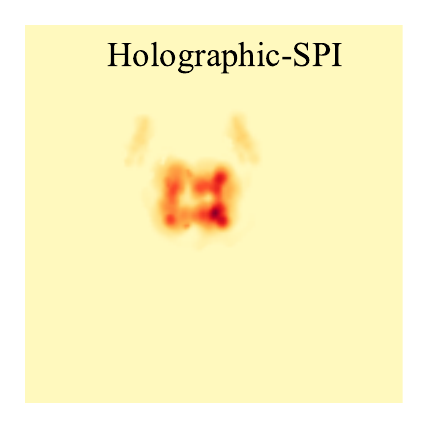}&\includegraphics[scale=0.67]{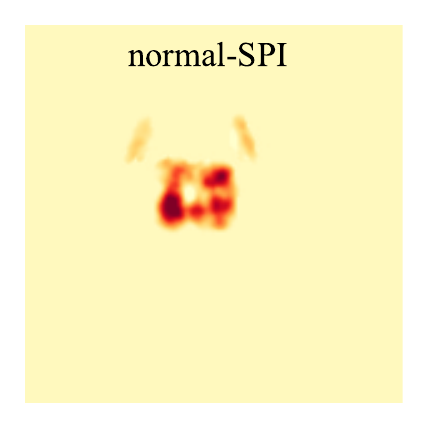} \\
(a) & (b) & (c)\\[6pt]
\includegraphics[scale=0.67]{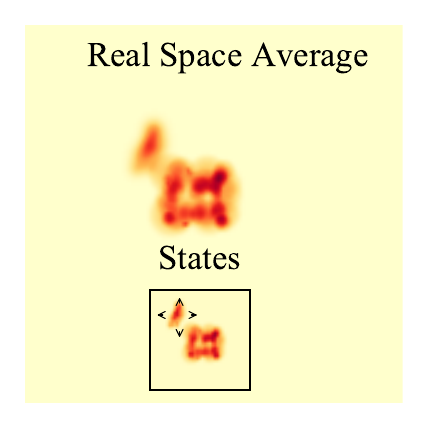}&\includegraphics[scale=0.67]{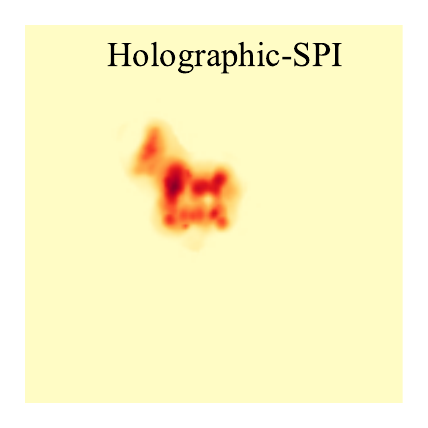}&\includegraphics[scale=0.67]{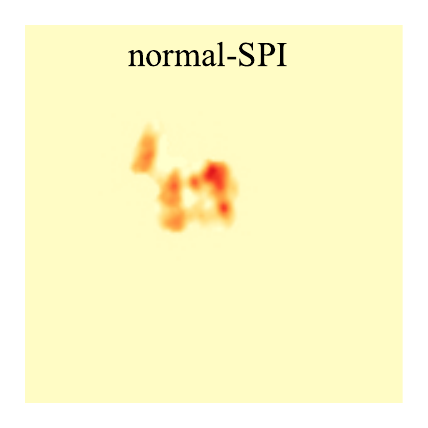} \\
(d) & (e) & (f)\\[6pt]
\end{tabular}
\begin{tabular}{c}
\includegraphics[scale=0.67]{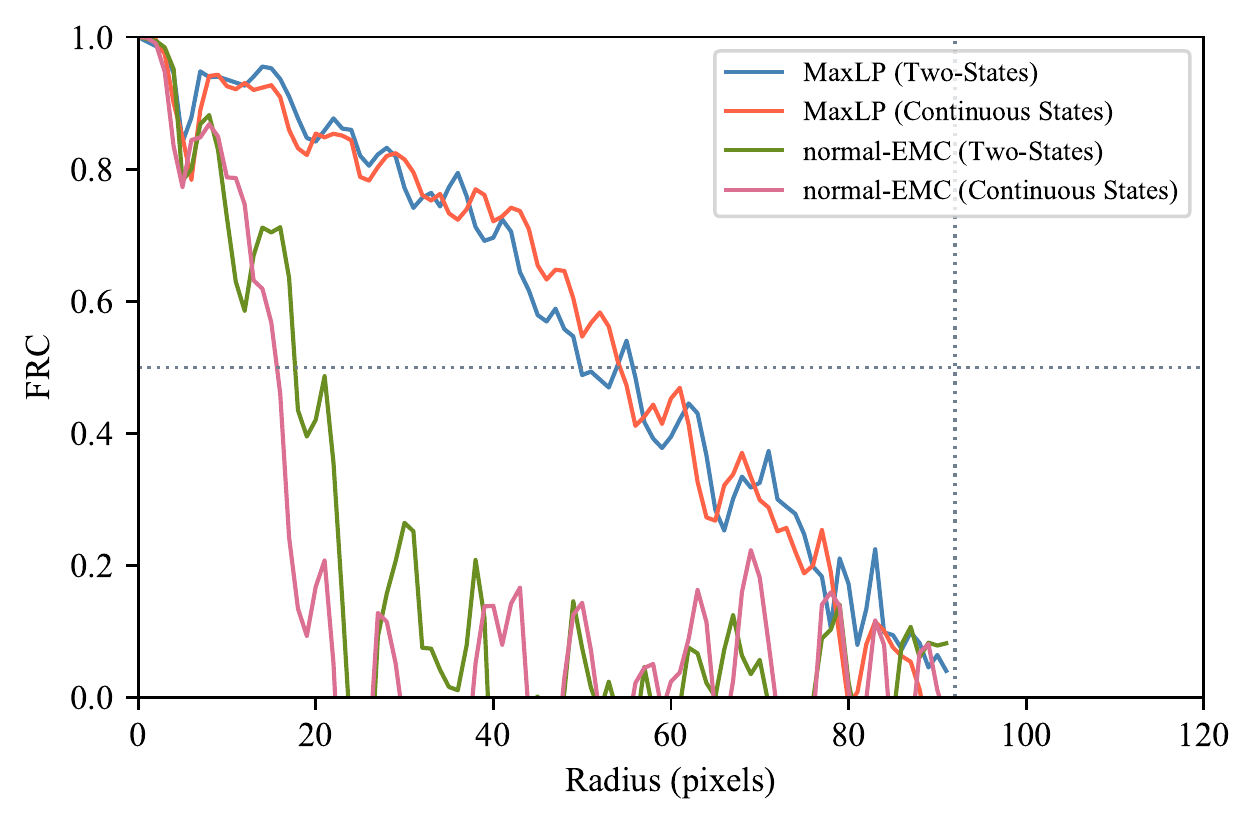}\\
(g)\\[6pt]
\end{tabular}\\
\caption{Simulation results for structural heterogeneity in the target object (a) Average structure of target object which exists in one of two discrete states (State A $\&$ State B, shown in inset). (b) Reconstructed real-space target object with an AuNP sphere attached using MaxLP. (c) Reconstructed real-space target object without reference attached. (d) Average structure of target object which exists in a continuous distribution of states. Inset shows the target object with random sub-unit where arrow depicts the free direction of motion for sub-unit in each shot. (e) Reconstructed real-space target object with an AuNP sphere attached using MaxLP. (f) Reconstructed real-space target object without reference attached. (g) Comparison of Fourier ring correlation (FRC) between holographic-SPI and normal-SPI for the different scenarios.} 
\label{fig:hetro}
\end{figure}

Many biological entities have flexible sub-units that can move following a continuous landscape of various conformational states. The study of these conformation changes can help us understand the function of such biomolecules. However, there are significant computational challenges to retrieve conformational states and structure of target object from diffraction data. In conventional SPI, the averaging of patterns from variable structures has an undetermined effect on the retrieved structure. This is because even though the Fourier transform is a linear operator, it's the intensities (squared magnitudes) that are averaged across different frames. In general, the resultant intensity distribution has lower contrast and is not the Fourier transform of any compact object. In some cases with uncorrelated random motion of atoms, one could expect the background-subtracted intensities to represent the scattering from the average structure, but this is in no way guaranteed, especially with correlated motions. Thus, extensive computational efforts must be employed to discover a subset of data from a homogeneous ensemble or to solve for additional latent parameters associated with structural variations.

In the current holographic regime, we solve for the complex Fourier transform, which is linearly related to the real-space structure. Thus, we may expect the reconstruction to be that of the average structure. To examine whether this is indeed the case, we perform two computational experiments below where the target object varies from frame to frame but the algorithm still attempts to reconstruct a single structure.


\subsubsection{Two-state heterogeneity}

We investigate the scenario in which the target object is composed of a homogeneous rigid part (the large unit) and a moving sub-unit that exists in different states shot-to-shot. Without explicitly modifying the MaxLP algorithm or providing any information about such heterogeneity, the algorithm analyzes the diffraction dataset simulated from these target objects. In the first case, the moving sub-unit object exists in two distinct discrete states: State A \& State B, as shown in inset of Fig.~\ref{fig:hetro}(a). The larger sub-unit remains in same position, whereas the smaller sub-unit occupies one of two opposite states about the center of the larger one.

The diffraction patterns are now generated from this heterogeneous ensemble in two ways, one with the AuNP attached and the other with just the target object (as in conventional SPI). In the former case, we still have the shot-to-shot variations in the AuNP diameter and relative position as before, which are recovered with the MaxLP procedure developed above. For the latter case, conventional intensity reconstruction is performed with the EMC algorithm, followed by phase retrieval. 
Figure~\ref{fig:hetro}(b-c) shows the reconstructions from the holographic and normal SPI scenarios. In the holographic case, the reconstruction is an object very similar to the real-space average of structure shown in Fig.~\ref{fig:hetro}(a). However, in the absence of a reference, as shown in Fig.~\ref{fig:hetro}(c), the reconstruction after phase retrieval is quite poor. This is also seen in the FRC comparisons with the real-space average in Fig.~\ref{fig:hetro}(g). 

We would like to stress that this result, while expected, is not trivial. First, we are still measuring real-valued intensities and not complex Fourier amplitudes. Secondly, each pattern has a different set of latent variables, leading to different intensity distributions on the detector. The difference is that the MaxLP algorithm does not average these intensities directly, but rather fits a common complex Fourier transform of the target object to all these intensities, which turns out to reflect the average object over all patterns.

\subsubsection{Continuous heterogeneity}
The previous example addressed the extreme case of having the moving sub-unit in two diametrically opposite locations. We now investigate a more realistic scenario where the moving sub-unit occupies a continuous local distribution of positions. The displacements are sampled from a normal distribution with $\sigma=0.5$ pixels, as represented by arrows in the inset of Fig.~\ref{fig:hetro}(d). The dataset consists of diffraction patterns generated from the target object where the smaller sub-unit is in a different position every frame. Figure~\ref{fig:hetro}(e-f) show the MaxLP and normal SPI reconstructions as before. In the holographic case, the homogeneous part is reconstructed to a good resolution while the wobbling sub-unit part has been reconstructed as a blurry object, as expected in a real-space average. This is in contrast to the normal SPI case, where the resolution of the reconstruction is globally affected due to the averaging in intensity space, leading to loss of contrast at higher $q$. Once again, the Fig.~\ref{fig:hetro}(g) shows that the holographic SPI reconstruction compares much more favorably with the average structure.

The critical aspect of this robustness to heterogeneity seems to be the presence of translational latent parameters. For the continuous translation case, one can envision sharpening the small sub-unit reconstruction by choosing the AuNP shifts relative to that sub-unit rather than whole object. Such an option is not possible in conventional SPI where the diffraction pattern is invariant to translating the object within the X-ray focus.

\section{\label{sec:concl}Conclusions}

In conventional SPI experiments, the biggest challenge in reaching sub-nm resolutions is to collect a large number of diffraction patterns with sufficiently low background in order to enable orientation determination and averaging. The introduced holography-based imaging methodology overcomes this issue by attaching a strong scattering object (a reference) to the sample of interest. this significantly improves background tolerance, but adds a large amount of computational complexity by introducing more unknown parameters associated with each pattern defining the conjugate system. In this paper, we described an algorithm based on maximum-likelihood estimation using pattern search called MaxLP that enables one to scale the method to high resolution and to weakly scattering objects. The results above show how it significantly outperforms the previously published method in the cases of low signal and without the need for a large number of intermediate average intensity models.

We extended the application of MaxLP to retrieve the structure of the target object for the scenario where it has shot-to-shot structural heterogeneity. Two cases of heterogeneity were investigated: one in which a sub-unit is in two discrete conformational states and one in which the sub-unit is in a continuous distribution of states. In both the cases, the algorithm reconstructs an object that is real space average of structure in all the conformation states, without prior knowledge that the target object was varying from shot-to-shot. 
On the contrary, the conventional SPI without an added holographic reference generates an average over the intensities in Fourier space that on phase retrieval yields a poorly reconstructed object not just in the vicinity of the moving sub-unit, but globally due to the delocalized nature of the Fourier transform. 

The algorithm can also be applied when the reference is a 2D crystal lattice and the target object is located in one of the unit cells. In such scenarios, one solves for relative shifts between the center of unit cell and target object in each shot and the size of unit cell. 

In future work, we will explore the possibility of adjusting the $xy$ shifts in order to selectively reconstruct different regions of the particle to varying degrees of sharpness. We also plan to explicitly incorporate structural heterogeneity either as multiple discrete classes~\cite{Ayyer:2021} or as continuous latent variables~\cite{Zhuang:2022}. While intensity-space classifications have been performed previously, the holographic approach may enable a more fine-grained approach similar to that applied in cryogenic electron microscopy~\cite{Zhong:2021}.

The data generation and algorithm codes used in this work are open access and are available at Github. 


\section{Acknowledgments}
The author acknowledges the extremely valuable discussions with Tamme Wollweber, Yulong Zhuang, Zhou Shen and Parichita Mazumder. This research is funded by Max Planck Society.

\section{Disclosures}
The authors declare no conflict of interest.

\bibliography{refs}

\begin{thebibliography}{31}%
\makeatletter
\providecommand \@ifxundefined [1]{%
 \@ifx{#1\undefined}
}%
\providecommand \@ifnum [1]{%
 \ifnum #1\expandafter \@firstoftwo
 \else \expandafter \@secondoftwo
 \fi
}%
\providecommand \@ifx [1]{%
 \ifx #1\expandafter \@firstoftwo
 \else \expandafter \@secondoftwo
 \fi
}%
\providecommand \natexlab [1]{#1}%
\providecommand \enquote  [1]{``#1''}%
\providecommand \bibnamefont  [1]{#1}%
\providecommand \bibfnamefont [1]{#1}%
\providecommand \citenamefont [1]{#1}%
\providecommand \href@noop [0]{\@secondoftwo}%
\providecommand \href [0]{\begingroup \@sanitize@url \@href}%
\providecommand \@href[1]{\@@startlink{#1}\@@href}%
\providecommand \@@href[1]{\endgroup#1\@@endlink}%
\providecommand \@sanitize@url [0]{\catcode `\\12\catcode `\$12\catcode
  `\&12\catcode `\#12\catcode `\^12\catcode `\_12\catcode `\%12\relax}%
\providecommand \@@startlink[1]{}%
\providecommand \@@endlink[0]{}%
\providecommand \url  [0]{\begingroup\@sanitize@url \@url }%
\providecommand \@url [1]{\endgroup\@href {#1}{\urlprefix }}%
\providecommand \urlprefix  [0]{URL }%
\providecommand \Eprint [0]{\href }%
\providecommand \doibase [0]{http://dx.doi.org/}%
\providecommand \selectlanguage [0]{\@gobble}%
\providecommand \bibinfo  [0]{\@secondoftwo}%
\providecommand \bibfield  [0]{\@secondoftwo}%
\providecommand \translation [1]{[#1]}%
\providecommand \BibitemOpen [0]{}%
\providecommand \bibitemStop [0]{}%
\providecommand \bibitemNoStop [0]{.\EOS\space}%
\providecommand \EOS [0]{\spacefactor3000\relax}%
\providecommand \BibitemShut  [1]{\csname bibitem#1\endcsname}%
\let\auto@bib@innerbib\@empty
\bibitem [{\citenamefont {Aquila}\ \emph {et~al.}(2015)\citenamefont {Aquila},
  \citenamefont {Barty}, \citenamefont {Bostedt}, \citenamefont {Boutet},
  \citenamefont {Carini}, \citenamefont {DePonte}, \citenamefont {Drell},
  \citenamefont {Doniach}, \citenamefont {Downing}, \citenamefont {Earnest}
  \emph {et~al.}}]{Aquila:2015}%
  \BibitemOpen
  \bibfield  {author} {\bibinfo {author} {\bibfnamefont {A.}~\bibnamefont
  {Aquila}}, \bibinfo {author} {\bibfnamefont {A.}~\bibnamefont {Barty}},
  \bibinfo {author} {\bibfnamefont {C.}~\bibnamefont {Bostedt}}, \bibinfo
  {author} {\bibfnamefont {S.}~\bibnamefont {Boutet}}, \bibinfo {author}
  {\bibfnamefont {G.}~\bibnamefont {Carini}}, \bibinfo {author} {\bibfnamefont
  {D.}~\bibnamefont {DePonte}}, \bibinfo {author} {\bibfnamefont
  {P.}~\bibnamefont {Drell}}, \bibinfo {author} {\bibfnamefont
  {S.}~\bibnamefont {Doniach}}, \bibinfo {author} {\bibfnamefont
  {K.}~\bibnamefont {Downing}}, \bibinfo {author} {\bibfnamefont
  {T.}~\bibnamefont {Earnest}},  \emph {et~al.},\ }\href@noop {} {\bibfield
  {journal} {\bibinfo  {journal} {Structural Dynamics}\ }\textbf {\bibinfo
  {volume} {2}},\ \bibinfo {pages} {041701} (\bibinfo {year}
  {2015})}\BibitemShut {NoStop}%
\bibitem [{\citenamefont {Chapman}\ \emph {et~al.}(2006)\citenamefont
  {Chapman}, \citenamefont {Barty}, \citenamefont {Bogan}, \citenamefont
  {Boutet}, \citenamefont {Frank}, \citenamefont {Hau-Riege}, \citenamefont
  {Marchesini}, \citenamefont {Woods}, \citenamefont {Bajt}, \citenamefont
  {Benner} \emph {et~al.}}]{Chapman:2006}%
  \BibitemOpen
  \bibfield  {author} {\bibinfo {author} {\bibfnamefont {H.~N.}\ \bibnamefont
  {Chapman}}, \bibinfo {author} {\bibfnamefont {A.}~\bibnamefont {Barty}},
  \bibinfo {author} {\bibfnamefont {M.~J.}\ \bibnamefont {Bogan}}, \bibinfo
  {author} {\bibfnamefont {S.}~\bibnamefont {Boutet}}, \bibinfo {author}
  {\bibfnamefont {M.}~\bibnamefont {Frank}}, \bibinfo {author} {\bibfnamefont
  {S.~P.}\ \bibnamefont {Hau-Riege}}, \bibinfo {author} {\bibfnamefont
  {S.}~\bibnamefont {Marchesini}}, \bibinfo {author} {\bibfnamefont {B.~W.}\
  \bibnamefont {Woods}}, \bibinfo {author} {\bibfnamefont {S.}~\bibnamefont
  {Bajt}}, \bibinfo {author} {\bibfnamefont {W.~H.}\ \bibnamefont {Benner}},
  \emph {et~al.},\ }\href {\doibase 10.1038/nphys461} {\bibfield  {journal}
  {\bibinfo  {journal} {Nature Physics}\ }\textbf {\bibinfo {volume} {2}},\
  \bibinfo {pages} {839} (\bibinfo {year} {2006})}\BibitemShut {NoStop}%
\bibitem [{\citenamefont {Sobolev}\ \emph {et~al.}(2020)\citenamefont
  {Sobolev}, \citenamefont {Zolotarev}, \citenamefont {Giewekemeyer},
  \citenamefont {Bielecki}, \citenamefont {Okamoto}, \citenamefont {Reddy},
  \citenamefont {Andreasson}, \citenamefont {Ayyer}, \citenamefont {Barak},
  \citenamefont {Bari} \emph {et~al.}}]{Sobolev:2020}%
  \BibitemOpen
  \bibfield  {author} {\bibinfo {author} {\bibfnamefont {E.}~\bibnamefont
  {Sobolev}}, \bibinfo {author} {\bibfnamefont {S.}~\bibnamefont {Zolotarev}},
  \bibinfo {author} {\bibfnamefont {K.}~\bibnamefont {Giewekemeyer}}, \bibinfo
  {author} {\bibfnamefont {J.}~\bibnamefont {Bielecki}}, \bibinfo {author}
  {\bibfnamefont {K.}~\bibnamefont {Okamoto}}, \bibinfo {author} {\bibfnamefont
  {H.~K.}\ \bibnamefont {Reddy}}, \bibinfo {author} {\bibfnamefont
  {J.}~\bibnamefont {Andreasson}}, \bibinfo {author} {\bibfnamefont
  {K.}~\bibnamefont {Ayyer}}, \bibinfo {author} {\bibfnamefont
  {I.}~\bibnamefont {Barak}}, \bibinfo {author} {\bibfnamefont
  {S.}~\bibnamefont {Bari}},  \emph {et~al.},\ }\href@noop {} {\bibfield
  {journal} {\bibinfo  {journal} {Communications Physics}\ }\textbf {\bibinfo
  {volume} {3}},\ \bibinfo {pages} {1} (\bibinfo {year} {2020})}\BibitemShut
  {NoStop}%
\bibitem [{\citenamefont {Yoon}\ \emph {et~al.}(2011)\citenamefont {Yoon},
  \citenamefont {Schwander}, \citenamefont {Abergel}, \citenamefont
  {Andersson}, \citenamefont {Andreasson}, \citenamefont {Aquila},
  \citenamefont {Bajt}, \citenamefont {Barthelmess}, \citenamefont {Barty},
  \citenamefont {Bogan} \emph {et~al.}}]{Yoon:2011}%
  \BibitemOpen
  \bibfield  {author} {\bibinfo {author} {\bibfnamefont {C.~H.}\ \bibnamefont
  {Yoon}}, \bibinfo {author} {\bibfnamefont {P.}~\bibnamefont {Schwander}},
  \bibinfo {author} {\bibfnamefont {C.}~\bibnamefont {Abergel}}, \bibinfo
  {author} {\bibfnamefont {I.}~\bibnamefont {Andersson}}, \bibinfo {author}
  {\bibfnamefont {J.}~\bibnamefont {Andreasson}}, \bibinfo {author}
  {\bibfnamefont {A.}~\bibnamefont {Aquila}}, \bibinfo {author} {\bibfnamefont
  {S.}~\bibnamefont {Bajt}}, \bibinfo {author} {\bibfnamefont {M.}~\bibnamefont
  {Barthelmess}}, \bibinfo {author} {\bibfnamefont {A.}~\bibnamefont {Barty}},
  \bibinfo {author} {\bibfnamefont {M.~J.}\ \bibnamefont {Bogan}},  \emph
  {et~al.},\ }\href {\doibase 10.1364/OE.19.016542} {\bibfield  {journal}
  {\bibinfo  {journal} {Opt. Express}\ }\textbf {\bibinfo {volume} {19}},\
  \bibinfo {pages} {16542} (\bibinfo {year} {2011})}\BibitemShut {NoStop}%
\bibitem [{\citenamefont {Ayyer}\ \emph {et~al.}(2021)\citenamefont {Ayyer},
  \citenamefont {Xavier}, \citenamefont {Bielecki}, \citenamefont {Shen},
  \citenamefont {Daurer}, \citenamefont {Samanta}, \citenamefont {Awel},
  \citenamefont {Bean}, \citenamefont {Barty}, \citenamefont {Bergemann} \emph
  {et~al.}}]{Ayyer:2021}%
  \BibitemOpen
  \bibfield  {author} {\bibinfo {author} {\bibfnamefont {K.}~\bibnamefont
  {Ayyer}}, \bibinfo {author} {\bibfnamefont {P.~L.}\ \bibnamefont {Xavier}},
  \bibinfo {author} {\bibfnamefont {J.}~\bibnamefont {Bielecki}}, \bibinfo
  {author} {\bibfnamefont {Z.}~\bibnamefont {Shen}}, \bibinfo {author}
  {\bibfnamefont {B.~J.}\ \bibnamefont {Daurer}}, \bibinfo {author}
  {\bibfnamefont {A.~K.}\ \bibnamefont {Samanta}}, \bibinfo {author}
  {\bibfnamefont {S.}~\bibnamefont {Awel}}, \bibinfo {author} {\bibfnamefont
  {R.}~\bibnamefont {Bean}}, \bibinfo {author} {\bibfnamefont {A.}~\bibnamefont
  {Barty}}, \bibinfo {author} {\bibfnamefont {M.}~\bibnamefont {Bergemann}},
  \emph {et~al.},\ }\href {\doibase 10.1364/OPTICA.410851} {\bibfield
  {journal} {\bibinfo  {journal} {Optica}\ }\textbf {\bibinfo {volume} {8}},\
  \bibinfo {pages} {15} (\bibinfo {year} {2021})}\BibitemShut {NoStop}%
\bibitem [{\citenamefont {Loh}\ and\ \citenamefont {Elser}(2009)}]{Loh:2009}%
  \BibitemOpen
  \bibfield  {author} {\bibinfo {author} {\bibfnamefont {N.-T.~D.}\
  \bibnamefont {Loh}}\ and\ \bibinfo {author} {\bibfnamefont {V.}~\bibnamefont
  {Elser}},\ }\href@noop {} {\bibfield  {journal} {\bibinfo  {journal}
  {Physical Review E}\ }\textbf {\bibinfo {volume} {80}},\ \bibinfo {pages}
  {026705} (\bibinfo {year} {2009})}\BibitemShut {NoStop}%
\bibitem [{\citenamefont {Ayyer}\ \emph {et~al.}(2016)\citenamefont {Ayyer},
  \citenamefont {Lan}, \citenamefont {Elser},\ and\ \citenamefont
  {Loh}}]{Ayyer:2016}%
  \BibitemOpen
  \bibfield  {author} {\bibinfo {author} {\bibfnamefont {K.}~\bibnamefont
  {Ayyer}}, \bibinfo {author} {\bibfnamefont {T.-Y.}\ \bibnamefont {Lan}},
  \bibinfo {author} {\bibfnamefont {V.}~\bibnamefont {Elser}}, \ and\ \bibinfo
  {author} {\bibfnamefont {N.~D.}\ \bibnamefont {Loh}},\ }\href@noop {}
  {\bibfield  {journal} {\bibinfo  {journal} {Journal of applied
  crystallography}\ }\textbf {\bibinfo {volume} {49}},\ \bibinfo {pages} {1320}
  (\bibinfo {year} {2016})}\BibitemShut {NoStop}%
\bibitem [{\citenamefont {Fienup}(1978)}]{Fienup:1978}%
  \BibitemOpen
  \bibfield  {author} {\bibinfo {author} {\bibfnamefont {J.~R.}\ \bibnamefont
  {Fienup}},\ }\href@noop {} {\bibfield  {journal} {\bibinfo  {journal} {Optics
  letters}\ }\textbf {\bibinfo {volume} {3}},\ \bibinfo {pages} {27} (\bibinfo
  {year} {1978})}\BibitemShut {NoStop}%
\bibitem [{\citenamefont {Elser}(2003)}]{Elser:2003}%
  \BibitemOpen
  \bibfield  {author} {\bibinfo {author} {\bibfnamefont {V.}~\bibnamefont
  {Elser}},\ }\href@noop {} {\bibfield  {journal} {\bibinfo  {journal} {JOSA
  A}\ }\textbf {\bibinfo {volume} {20}},\ \bibinfo {pages} {40} (\bibinfo
  {year} {2003})}\BibitemShut {NoStop}%
\bibitem [{\citenamefont {Lundholm}\ \emph {et~al.}(2018)\citenamefont
  {Lundholm}, \citenamefont {Sellberg}, \citenamefont {Ekeberg}, \citenamefont
  {Hantke}, \citenamefont {Okamoto}, \citenamefont {van~der Schot},
  \citenamefont {Andreasson}, \citenamefont {Barty}, \citenamefont {Bielecki},
  \citenamefont {Bruza} \emph {et~al.}}]{Lundholm:2018}%
  \BibitemOpen
  \bibfield  {author} {\bibinfo {author} {\bibfnamefont {I.~V.}\ \bibnamefont
  {Lundholm}}, \bibinfo {author} {\bibfnamefont {J.~A.}\ \bibnamefont
  {Sellberg}}, \bibinfo {author} {\bibfnamefont {T.}~\bibnamefont {Ekeberg}},
  \bibinfo {author} {\bibfnamefont {M.~F.}\ \bibnamefont {Hantke}}, \bibinfo
  {author} {\bibfnamefont {K.}~\bibnamefont {Okamoto}}, \bibinfo {author}
  {\bibfnamefont {G.}~\bibnamefont {van~der Schot}}, \bibinfo {author}
  {\bibfnamefont {J.}~\bibnamefont {Andreasson}}, \bibinfo {author}
  {\bibfnamefont {A.}~\bibnamefont {Barty}}, \bibinfo {author} {\bibfnamefont
  {J.}~\bibnamefont {Bielecki}}, \bibinfo {author} {\bibfnamefont
  {P.}~\bibnamefont {Bruza}},  \emph {et~al.},\ }\href {\doibase
  10.1107/S2052252518010047} {\bibfield  {journal} {\bibinfo  {journal}
  {IUCrJ}\ }\textbf {\bibinfo {volume} {5}},\ \bibinfo {pages} {531} (\bibinfo
  {year} {2018})}\BibitemShut {NoStop}%
\bibitem [{\citenamefont {Ayyer}\ \emph {et~al.}(2019)\citenamefont {Ayyer},
  \citenamefont {Morgan}, \citenamefont {Aquila}, \citenamefont {DeMirci},
  \citenamefont {Hogue}, \citenamefont {Kirian}, \citenamefont {Xavier},
  \citenamefont {Yoon}, \citenamefont {Chapman},\ and\ \citenamefont
  {Barty}}]{Ayyer:2019}%
  \BibitemOpen
  \bibfield  {author} {\bibinfo {author} {\bibfnamefont {K.}~\bibnamefont
  {Ayyer}}, \bibinfo {author} {\bibfnamefont {A.~J.}\ \bibnamefont {Morgan}},
  \bibinfo {author} {\bibfnamefont {A.}~\bibnamefont {Aquila}}, \bibinfo
  {author} {\bibfnamefont {H.}~\bibnamefont {DeMirci}}, \bibinfo {author}
  {\bibfnamefont {B.~G.}\ \bibnamefont {Hogue}}, \bibinfo {author}
  {\bibfnamefont {R.~A.}\ \bibnamefont {Kirian}}, \bibinfo {author}
  {\bibfnamefont {P.~L.}\ \bibnamefont {Xavier}}, \bibinfo {author}
  {\bibfnamefont {C.~H.}\ \bibnamefont {Yoon}}, \bibinfo {author}
  {\bibfnamefont {H.~N.}\ \bibnamefont {Chapman}}, \ and\ \bibinfo {author}
  {\bibfnamefont {A.}~\bibnamefont {Barty}},\ }\href@noop {} {\bibfield
  {journal} {\bibinfo  {journal} {Optics Express}\ }\textbf {\bibinfo {volume}
  {27}},\ \bibinfo {pages} {37816} (\bibinfo {year} {2019})}\BibitemShut
  {NoStop}%
\bibitem [{\citenamefont {Ekeberg}\ \emph {et~al.}(2015)\citenamefont
  {Ekeberg}, \citenamefont {Svenda}, \citenamefont {Abergel}, \citenamefont
  {Maia}, \citenamefont {Seltzer}, \citenamefont {Claverie}, \citenamefont
  {Hantke}, \citenamefont {J\"onsson}, \citenamefont {Nettelblad},
  \citenamefont {van~der Schot} \emph {et~al.}}]{Ekeberg:2015}%
  \BibitemOpen
  \bibfield  {author} {\bibinfo {author} {\bibfnamefont {T.}~\bibnamefont
  {Ekeberg}}, \bibinfo {author} {\bibfnamefont {M.}~\bibnamefont {Svenda}},
  \bibinfo {author} {\bibfnamefont {C.}~\bibnamefont {Abergel}}, \bibinfo
  {author} {\bibfnamefont {F.~R. N.~C.}\ \bibnamefont {Maia}}, \bibinfo
  {author} {\bibfnamefont {V.}~\bibnamefont {Seltzer}}, \bibinfo {author}
  {\bibfnamefont {J.-M.}\ \bibnamefont {Claverie}}, \bibinfo {author}
  {\bibfnamefont {M.}~\bibnamefont {Hantke}}, \bibinfo {author} {\bibfnamefont
  {O.}~\bibnamefont {J\"onsson}}, \bibinfo {author} {\bibfnamefont
  {C.}~\bibnamefont {Nettelblad}}, \bibinfo {author} {\bibfnamefont
  {G.}~\bibnamefont {van~der Schot}},  \emph {et~al.},\ }\href {\doibase
  10.1103/PhysRevLett.114.098102} {\bibfield  {journal} {\bibinfo  {journal}
  {Phys. Rev. Lett.}\ }\textbf {\bibinfo {volume} {114}},\ \bibinfo {pages}
  {098102} (\bibinfo {year} {2015})}\BibitemShut {NoStop}%
\bibitem [{\citenamefont {Rose}\ \emph {et~al.}(2018)\citenamefont {Rose},
  \citenamefont {Bobkov}, \citenamefont {Ayyer}, \citenamefont {Kurta},
  \citenamefont {Dzhigaev}, \citenamefont {Kim}, \citenamefont {Morgan},
  \citenamefont {Yoon}, \citenamefont {Westphal}, \citenamefont {Bielecki}
  \emph {et~al.}}]{Rose:2018}%
  \BibitemOpen
  \bibfield  {author} {\bibinfo {author} {\bibfnamefont {M.}~\bibnamefont
  {Rose}}, \bibinfo {author} {\bibfnamefont {S.}~\bibnamefont {Bobkov}},
  \bibinfo {author} {\bibfnamefont {K.}~\bibnamefont {Ayyer}}, \bibinfo
  {author} {\bibfnamefont {R.~P.}\ \bibnamefont {Kurta}}, \bibinfo {author}
  {\bibfnamefont {D.}~\bibnamefont {Dzhigaev}}, \bibinfo {author}
  {\bibfnamefont {Y.~Y.}\ \bibnamefont {Kim}}, \bibinfo {author} {\bibfnamefont
  {A.~J.}\ \bibnamefont {Morgan}}, \bibinfo {author} {\bibfnamefont {C.~H.}\
  \bibnamefont {Yoon}}, \bibinfo {author} {\bibfnamefont {D.}~\bibnamefont
  {Westphal}}, \bibinfo {author} {\bibfnamefont {J.}~\bibnamefont {Bielecki}},
  \emph {et~al.},\ }\href@noop {} {\bibfield  {journal} {\bibinfo  {journal}
  {IUCrJ}\ }\textbf {\bibinfo {volume} {5}} (\bibinfo {year}
  {2018})}\BibitemShut {NoStop}%
\bibitem [{\citenamefont {Philipp}\ \emph {et~al.}(2012)\citenamefont
  {Philipp}, \citenamefont {Ayyer}, \citenamefont {Tate}, \citenamefont
  {Elser},\ and\ \citenamefont {Gruner}}]{Philipp:2012}%
  \BibitemOpen
  \bibfield  {author} {\bibinfo {author} {\bibfnamefont {H.~T.}\ \bibnamefont
  {Philipp}}, \bibinfo {author} {\bibfnamefont {K.}~\bibnamefont {Ayyer}},
  \bibinfo {author} {\bibfnamefont {M.~W.}\ \bibnamefont {Tate}}, \bibinfo
  {author} {\bibfnamefont {V.}~\bibnamefont {Elser}}, \ and\ \bibinfo {author}
  {\bibfnamefont {S.~M.}\ \bibnamefont {Gruner}},\ }\href@noop {} {\bibfield
  {journal} {\bibinfo  {journal} {Optics express}\ }\textbf {\bibinfo {volume}
  {20}},\ \bibinfo {pages} {13129} (\bibinfo {year} {2012})}\BibitemShut
  {NoStop}%
\bibitem [{\citenamefont {Poudyal}\ \emph {et~al.}(2020)\citenamefont
  {Poudyal}, \citenamefont {Schmidt},\ and\ \citenamefont
  {Schwander}}]{Poudyal:2020}%
  \BibitemOpen
  \bibfield  {author} {\bibinfo {author} {\bibfnamefont {I.}~\bibnamefont
  {Poudyal}}, \bibinfo {author} {\bibfnamefont {M.}~\bibnamefont {Schmidt}}, \
  and\ \bibinfo {author} {\bibfnamefont {P.}~\bibnamefont {Schwander}},\
  }\href@noop {} {\bibfield  {journal} {\bibinfo  {journal} {Structural
  Dynamics}\ }\textbf {\bibinfo {volume} {7}},\ \bibinfo {pages} {024102}
  (\bibinfo {year} {2020})}\BibitemShut {NoStop}%
\bibitem [{\citenamefont {Munke}\ \emph {et~al.}(2016)\citenamefont {Munke},
  \citenamefont {Andreasson}, \citenamefont {Aquila}, \citenamefont {Awel},
  \citenamefont {Ayyer}, \citenamefont {Barty}, \citenamefont {Bean},
  \citenamefont {Berntsen}, \citenamefont {Bielecki}, \citenamefont {Boutet}
  \emph {et~al.}}]{Munke:2016}%
  \BibitemOpen
  \bibfield  {author} {\bibinfo {author} {\bibfnamefont {A.}~\bibnamefont
  {Munke}}, \bibinfo {author} {\bibfnamefont {J.}~\bibnamefont {Andreasson}},
  \bibinfo {author} {\bibfnamefont {A.}~\bibnamefont {Aquila}}, \bibinfo
  {author} {\bibfnamefont {S.}~\bibnamefont {Awel}}, \bibinfo {author}
  {\bibfnamefont {K.}~\bibnamefont {Ayyer}}, \bibinfo {author} {\bibfnamefont
  {A.}~\bibnamefont {Barty}}, \bibinfo {author} {\bibfnamefont {R.~J.}\
  \bibnamefont {Bean}}, \bibinfo {author} {\bibfnamefont {P.}~\bibnamefont
  {Berntsen}}, \bibinfo {author} {\bibfnamefont {J.}~\bibnamefont {Bielecki}},
  \bibinfo {author} {\bibfnamefont {S.}~\bibnamefont {Boutet}},  \emph
  {et~al.},\ }\href {https://doi.org/10.1038/sdata.2016.64} {\bibfield
  {journal} {\bibinfo  {journal} {Scientific Data}\ }\textbf {\bibinfo {volume}
  {3}},\ \bibinfo {pages} {160064} (\bibinfo {year} {2016})},\ \bibinfo {note}
  {data Descriptor}\BibitemShut {NoStop}%
\bibitem [{\citenamefont {Bielecki}\ \emph {et~al.}(2019)\citenamefont
  {Bielecki}, \citenamefont {Hantke}, \citenamefont {Daurer}, \citenamefont
  {Reddy}, \citenamefont {Hasse}, \citenamefont {Larsson}, \citenamefont
  {Gunn}, \citenamefont {Svenda}, \citenamefont {Munke}, \citenamefont
  {Sellberg} \emph {et~al.}}]{Bielecki:2019}%
  \BibitemOpen
  \bibfield  {author} {\bibinfo {author} {\bibfnamefont {J.}~\bibnamefont
  {Bielecki}}, \bibinfo {author} {\bibfnamefont {M.~F.}\ \bibnamefont
  {Hantke}}, \bibinfo {author} {\bibfnamefont {B.~J.}\ \bibnamefont {Daurer}},
  \bibinfo {author} {\bibfnamefont {H.~K.~N.}\ \bibnamefont {Reddy}}, \bibinfo
  {author} {\bibfnamefont {D.}~\bibnamefont {Hasse}}, \bibinfo {author}
  {\bibfnamefont {D.~S.~D.}\ \bibnamefont {Larsson}}, \bibinfo {author}
  {\bibfnamefont {L.~H.}\ \bibnamefont {Gunn}}, \bibinfo {author}
  {\bibfnamefont {M.}~\bibnamefont {Svenda}}, \bibinfo {author} {\bibfnamefont
  {A.}~\bibnamefont {Munke}}, \bibinfo {author} {\bibfnamefont {J.~A.}\
  \bibnamefont {Sellberg}},  \emph {et~al.},\ }\href {\doibase
  10.1126/sciadv.aav8801} {\bibfield  {journal} {\bibinfo  {journal} {Science
  Advances}\ }\textbf {\bibinfo {volume} {5}} (\bibinfo {year} {2019}),\
  10.1126/sciadv.aav8801}\BibitemShut {NoStop}%
\bibitem [{\citenamefont {Chapman}\ \emph {et~al.}(2011)\citenamefont
  {Chapman}, \citenamefont {Fromme}, \citenamefont {Barty}, \citenamefont
  {White}, \citenamefont {Kirian}, \citenamefont {Aquila}, \citenamefont
  {Hunter}, \citenamefont {Schulz}, \citenamefont {DePonte}, \citenamefont
  {Weierstall} \emph {et~al.}}]{Chapman:2011}%
  \BibitemOpen
  \bibfield  {author} {\bibinfo {author} {\bibfnamefont {H.~N.}\ \bibnamefont
  {Chapman}}, \bibinfo {author} {\bibfnamefont {P.}~\bibnamefont {Fromme}},
  \bibinfo {author} {\bibfnamefont {A.}~\bibnamefont {Barty}}, \bibinfo
  {author} {\bibfnamefont {T.~A.}\ \bibnamefont {White}}, \bibinfo {author}
  {\bibfnamefont {R.~A.}\ \bibnamefont {Kirian}}, \bibinfo {author}
  {\bibfnamefont {A.}~\bibnamefont {Aquila}}, \bibinfo {author} {\bibfnamefont
  {M.~S.}\ \bibnamefont {Hunter}}, \bibinfo {author} {\bibfnamefont
  {J.}~\bibnamefont {Schulz}}, \bibinfo {author} {\bibfnamefont {D.~P.}\
  \bibnamefont {DePonte}}, \bibinfo {author} {\bibfnamefont {U.}~\bibnamefont
  {Weierstall}},  \emph {et~al.},\ }\href {\doibase 10.1038/nature09750}
  {\bibfield  {journal} {\bibinfo  {journal} {Nature}\ }\textbf {\bibinfo
  {volume} {470}},\ \bibinfo {pages} {73} (\bibinfo {year} {2011})}\BibitemShut
  {NoStop}%
\bibitem [{\citenamefont {Sierra}\ \emph {et~al.}(2012)\citenamefont {Sierra},
  \citenamefont {Laksmono}, \citenamefont {Kern}, \citenamefont {Tran},
  \citenamefont {Hattne}, \citenamefont {Alonso-Mori}, \citenamefont
  {Lassalle-Kaiser}, \citenamefont {Gl{\"{o}}ckner}, \citenamefont {Hellmich},
  \citenamefont {Schafer} \emph {et~al.}}]{Sierra:2012}%
  \BibitemOpen
  \bibfield  {author} {\bibinfo {author} {\bibfnamefont {R.~G.}\ \bibnamefont
  {Sierra}}, \bibinfo {author} {\bibfnamefont {H.}~\bibnamefont {Laksmono}},
  \bibinfo {author} {\bibfnamefont {J.}~\bibnamefont {Kern}}, \bibinfo {author}
  {\bibfnamefont {R.}~\bibnamefont {Tran}}, \bibinfo {author} {\bibfnamefont
  {J.}~\bibnamefont {Hattne}}, \bibinfo {author} {\bibfnamefont
  {R.}~\bibnamefont {Alonso-Mori}}, \bibinfo {author} {\bibfnamefont
  {B.}~\bibnamefont {Lassalle-Kaiser}}, \bibinfo {author} {\bibfnamefont
  {C.}~\bibnamefont {Gl{\"{o}}ckner}}, \bibinfo {author} {\bibfnamefont
  {J.}~\bibnamefont {Hellmich}}, \bibinfo {author} {\bibfnamefont {D.~W.}\
  \bibnamefont {Schafer}},  \emph {et~al.},\ }\href {\doibase
  10.1107/S0907444912038152} {\bibfield  {journal} {\bibinfo  {journal} {Acta
  Crystallographica Section D}\ }\textbf {\bibinfo {volume} {68}},\ \bibinfo
  {pages} {1584} (\bibinfo {year} {2012})}\BibitemShut {NoStop}%
\bibitem [{\citenamefont {Hunter}\ \emph {et~al.}(2014)\citenamefont {Hunter},
  \citenamefont {Segelke}, \citenamefont {Messerschmidt}, \citenamefont
  {Williams}, \citenamefont {Zatsepin}, \citenamefont {Barty}, \citenamefont
  {Benner}, \citenamefont {Carlson}, \citenamefont {Coleman}, \citenamefont
  {Graf} \emph {et~al.}}]{Hunter:2014}%
  \BibitemOpen
  \bibfield  {author} {\bibinfo {author} {\bibfnamefont {M.~S.}\ \bibnamefont
  {Hunter}}, \bibinfo {author} {\bibfnamefont {B.}~\bibnamefont {Segelke}},
  \bibinfo {author} {\bibfnamefont {M.}~\bibnamefont {Messerschmidt}}, \bibinfo
  {author} {\bibfnamefont {G.~J.}\ \bibnamefont {Williams}}, \bibinfo {author}
  {\bibfnamefont {N.~A.}\ \bibnamefont {Zatsepin}}, \bibinfo {author}
  {\bibfnamefont {A.}~\bibnamefont {Barty}}, \bibinfo {author} {\bibfnamefont
  {W.~H.}\ \bibnamefont {Benner}}, \bibinfo {author} {\bibfnamefont {D.~B.}\
  \bibnamefont {Carlson}}, \bibinfo {author} {\bibfnamefont {M.}~\bibnamefont
  {Coleman}}, \bibinfo {author} {\bibfnamefont {A.}~\bibnamefont {Graf}},
  \emph {et~al.},\ }\href@noop {} {\bibfield  {journal} {\bibinfo  {journal}
  {Scientific reports}\ }\textbf {\bibinfo {volume} {4}},\ \bibinfo {pages}
  {6026} (\bibinfo {year} {2014})}\BibitemShut {NoStop}%
\bibitem [{\citenamefont {Nam}\ \emph {et~al.}(2016)\citenamefont {Nam},
  \citenamefont {Kim}, \citenamefont {Kim}, \citenamefont {Ebisu},
  \citenamefont {Gallagher-Jones}, \citenamefont {Park}, \citenamefont {Kim},
  \citenamefont {Kim}, \citenamefont {Tono}, \citenamefont {Yabashi},
  \citenamefont {Ishikawa},\ and\ \citenamefont {Song}}]{Nam:2016}%
  \BibitemOpen
  \bibfield  {author} {\bibinfo {author} {\bibfnamefont {D.}~\bibnamefont
  {Nam}}, \bibinfo {author} {\bibfnamefont {C.}~\bibnamefont {Kim}}, \bibinfo
  {author} {\bibfnamefont {Y.}~\bibnamefont {Kim}}, \bibinfo {author}
  {\bibfnamefont {T.}~\bibnamefont {Ebisu}}, \bibinfo {author} {\bibfnamefont
  {M.}~\bibnamefont {Gallagher-Jones}}, \bibinfo {author} {\bibfnamefont
  {J.}~\bibnamefont {Park}}, \bibinfo {author} {\bibfnamefont {S.}~\bibnamefont
  {Kim}}, \bibinfo {author} {\bibfnamefont {S.}~\bibnamefont {Kim}}, \bibinfo
  {author} {\bibfnamefont {K.}~\bibnamefont {Tono}}, \bibinfo {author}
  {\bibfnamefont {M.}~\bibnamefont {Yabashi}}, \bibinfo {author} {\bibfnamefont
  {T.}~\bibnamefont {Ishikawa}}, \ and\ \bibinfo {author} {\bibfnamefont
  {C.}~\bibnamefont {Song}},\ }\href@noop {} {\bibfield  {journal} {\bibinfo
  {journal} {Journal of Physics B: Atomic, Molecular and Optical Physics}\
  }\textbf {\bibinfo {volume} {49}},\ \bibinfo {pages} {034008} (\bibinfo
  {year} {2016})}\BibitemShut {NoStop}%
\bibitem [{\citenamefont {Seuring}\ \emph {et~al.}(2018)\citenamefont
  {Seuring}, \citenamefont {Ayyer}, \citenamefont {Filippaki}, \citenamefont
  {Barthelmess}, \citenamefont {Longchamp}, \citenamefont {Ringler},
  \citenamefont {Pardini}, \citenamefont {Wojtas}, \citenamefont {Coleman},
  \citenamefont {D{\"o}rner} \emph {et~al.}}]{Seuring:2018}%
  \BibitemOpen
  \bibfield  {author} {\bibinfo {author} {\bibfnamefont {C.}~\bibnamefont
  {Seuring}}, \bibinfo {author} {\bibfnamefont {K.}~\bibnamefont {Ayyer}},
  \bibinfo {author} {\bibfnamefont {E.}~\bibnamefont {Filippaki}}, \bibinfo
  {author} {\bibfnamefont {M.}~\bibnamefont {Barthelmess}}, \bibinfo {author}
  {\bibfnamefont {J.-N.}\ \bibnamefont {Longchamp}}, \bibinfo {author}
  {\bibfnamefont {P.}~\bibnamefont {Ringler}}, \bibinfo {author} {\bibfnamefont
  {T.}~\bibnamefont {Pardini}}, \bibinfo {author} {\bibfnamefont {D.~H.}\
  \bibnamefont {Wojtas}}, \bibinfo {author} {\bibfnamefont {M.~A.}\
  \bibnamefont {Coleman}}, \bibinfo {author} {\bibfnamefont {K.}~\bibnamefont
  {D{\"o}rner}},  \emph {et~al.},\ }\href {\doibase 10.1038/s41467-018-04116-9}
  {\bibfield  {journal} {\bibinfo  {journal} {Nature Communications}\ }\textbf
  {\bibinfo {volume} {9}},\ \bibinfo {pages} {1836} (\bibinfo {year}
  {2018})}\BibitemShut {NoStop}%
\bibitem [{\citenamefont {Decking}\ \emph {et~al.}(2020)\citenamefont
  {Decking}, \citenamefont {Abeghyan}, \citenamefont {Abramian}, \citenamefont
  {Abramsky}, \citenamefont {Aguirre}, \citenamefont {Albrecht}, \citenamefont
  {Alou}, \citenamefont {Altarelli}, \citenamefont {Altmann}, \citenamefont
  {Amyan}, \citenamefont {Anashin}, \citenamefont {Mommerz}, \citenamefont
  {Monaco}, \citenamefont {Montiel}, \citenamefont {Moretti}, \citenamefont
  {Morozov}, \citenamefont {Morozov}, \citenamefont {Mross} \emph
  {et~al.}}]{Decking:2020}%
  \BibitemOpen
  \bibfield  {author} {\bibinfo {author} {\bibfnamefont {W.}~\bibnamefont
  {Decking}}, \bibinfo {author} {\bibfnamefont {S.}~\bibnamefont {Abeghyan}},
  \bibinfo {author} {\bibfnamefont {P.}~\bibnamefont {Abramian}}, \bibinfo
  {author} {\bibfnamefont {A.}~\bibnamefont {Abramsky}}, \bibinfo {author}
  {\bibfnamefont {A.}~\bibnamefont {Aguirre}}, \bibinfo {author} {\bibfnamefont
  {C.}~\bibnamefont {Albrecht}}, \bibinfo {author} {\bibfnamefont
  {P.}~\bibnamefont {Alou}}, \bibinfo {author} {\bibfnamefont {M.}~\bibnamefont
  {Altarelli}}, \bibinfo {author} {\bibfnamefont {P.}~\bibnamefont {Altmann}},
  \bibinfo {author} {\bibfnamefont {K.}~\bibnamefont {Amyan}}, \bibinfo
  {author} {\bibfnamefont {V.}~\bibnamefont {Anashin}}, \bibinfo {author}
  {\bibfnamefont {M.}~\bibnamefont {Mommerz}}, \bibinfo {author} {\bibfnamefont
  {L.}~\bibnamefont {Monaco}}, \bibinfo {author} {\bibfnamefont
  {C.}~\bibnamefont {Montiel}}, \bibinfo {author} {\bibfnamefont
  {M.}~\bibnamefont {Moretti}}, \bibinfo {author} {\bibfnamefont
  {I.}~\bibnamefont {Morozov}}, \bibinfo {author} {\bibfnamefont
  {P.}~\bibnamefont {Morozov}}, \bibinfo {author} {\bibfnamefont
  {D.}~\bibnamefont {Mross}},  \emph {et~al.},\ }\href {\doibase
  10.1038/s41566-020-0607-z} {\bibfield  {journal} {\bibinfo  {journal} {Nature
  Photonics}\ ,\ \bibinfo {pages} {391}} (\bibinfo {year} {2020})}\BibitemShut
  {NoStop}%
\bibitem [{\citenamefont {Ayyer}(2020)}]{Ayyer:2020}%
  \BibitemOpen
  \bibfield  {author} {\bibinfo {author} {\bibfnamefont {K.}~\bibnamefont
  {Ayyer}},\ }\href {\doibase 10.1364/OPTICA.391373} {\bibfield  {journal}
  {\bibinfo  {journal} {Optica}\ }\textbf {\bibinfo {volume} {7}},\ \bibinfo
  {pages} {593} (\bibinfo {year} {2020})}\BibitemShut {NoStop}%
\bibitem [{\citenamefont {Gravel}\ and\ \citenamefont
  {Elser}(2008)}]{Gravel:2008}%
  \BibitemOpen
  \bibfield  {author} {\bibinfo {author} {\bibfnamefont {S.}~\bibnamefont
  {Gravel}}\ and\ \bibinfo {author} {\bibfnamefont {V.}~\bibnamefont {Elser}},\
  }\href@noop {} {\bibfield  {journal} {\bibinfo  {journal} {Physical Review
  E}\ }\textbf {\bibinfo {volume} {78}},\ \bibinfo {pages} {036706} (\bibinfo
  {year} {2008})}\BibitemShut {NoStop}%
\bibitem [{\citenamefont {Pfeiffer}(2018)}]{Pfeiffer:2018}%
  \BibitemOpen
  \bibfield  {author} {\bibinfo {author} {\bibfnamefont {F.}~\bibnamefont
  {Pfeiffer}},\ }\href@noop {} {\bibfield  {journal} {\bibinfo  {journal}
  {Nature Photonics}\ }\textbf {\bibinfo {volume} {12}},\ \bibinfo {pages} {9}
  (\bibinfo {year} {2018})}\BibitemShut {NoStop}%
\bibitem [{\citenamefont {Thibault}\ and\ \citenamefont
  {Guizar-Sicairos}(2012)}]{Thibault:2012}%
  \BibitemOpen
  \bibfield  {author} {\bibinfo {author} {\bibfnamefont {P.}~\bibnamefont
  {Thibault}}\ and\ \bibinfo {author} {\bibfnamefont {M.}~\bibnamefont
  {Guizar-Sicairos}},\ }\href@noop {} {\bibfield  {journal} {\bibinfo
  {journal} {New Journal of Physics}\ }\textbf {\bibinfo {volume} {14}},\
  \bibinfo {pages} {063004} (\bibinfo {year} {2012})}\BibitemShut {NoStop}%
\bibitem [{\citenamefont {Torczon}(1997)}]{Torczon:1997}%
  \BibitemOpen
  \bibfield  {author} {\bibinfo {author} {\bibfnamefont {V.}~\bibnamefont
  {Torczon}},\ }\href@noop {} {\bibfield  {journal} {\bibinfo  {journal} {SIAM
  Journal on optimization}\ }\textbf {\bibinfo {volume} {7}},\ \bibinfo {pages}
  {1} (\bibinfo {year} {1997})}\BibitemShut {NoStop}%
\bibitem [{\citenamefont {Martin}\ \emph {et~al.}(2012)\citenamefont {Martin},
  \citenamefont {Loh}, \citenamefont {Hampton}, \citenamefont {Sierra},
  \citenamefont {Wang}, \citenamefont {Aquila}, \citenamefont {Bajt},
  \citenamefont {Barthelmess}, \citenamefont {Bostedt}, \citenamefont {Bozek}
  \emph {et~al.}}]{Martin:2012}%
  \BibitemOpen
  \bibfield  {author} {\bibinfo {author} {\bibfnamefont {A.~V.}\ \bibnamefont
  {Martin}}, \bibinfo {author} {\bibfnamefont {N.}~\bibnamefont {Loh}},
  \bibinfo {author} {\bibfnamefont {C.~Y.}\ \bibnamefont {Hampton}}, \bibinfo
  {author} {\bibfnamefont {R.~G.}\ \bibnamefont {Sierra}}, \bibinfo {author}
  {\bibfnamefont {F.}~\bibnamefont {Wang}}, \bibinfo {author} {\bibfnamefont
  {A.}~\bibnamefont {Aquila}}, \bibinfo {author} {\bibfnamefont
  {S.}~\bibnamefont {Bajt}}, \bibinfo {author} {\bibfnamefont {M.}~\bibnamefont
  {Barthelmess}}, \bibinfo {author} {\bibfnamefont {C.}~\bibnamefont
  {Bostedt}}, \bibinfo {author} {\bibfnamefont {J.~D.}\ \bibnamefont {Bozek}},
  \emph {et~al.},\ }\href@noop {} {\bibfield  {journal} {\bibinfo  {journal}
  {Optics express}\ }\textbf {\bibinfo {volume} {20}},\ \bibinfo {pages}
  {13501} (\bibinfo {year} {2012})}\BibitemShut {NoStop}%
\bibitem [{\citenamefont {Zhuang}\ \emph {et~al.}(2022)\citenamefont {Zhuang},
  \citenamefont {Awel}, \citenamefont {Barty}, \citenamefont {Bean},
  \citenamefont {Bielecki}, \citenamefont {Bergemann}, \citenamefont {Daurer},
  \citenamefont {Ekeberg}, \citenamefont {Estillore}, \citenamefont {Fangohr}
  \emph {et~al.}}]{Zhuang:2022}%
  \BibitemOpen
  \bibfield  {author} {\bibinfo {author} {\bibfnamefont {Y.}~\bibnamefont
  {Zhuang}}, \bibinfo {author} {\bibfnamefont {S.}~\bibnamefont {Awel}},
  \bibinfo {author} {\bibfnamefont {A.}~\bibnamefont {Barty}}, \bibinfo
  {author} {\bibfnamefont {R.}~\bibnamefont {Bean}}, \bibinfo {author}
  {\bibfnamefont {J.}~\bibnamefont {Bielecki}}, \bibinfo {author}
  {\bibfnamefont {M.}~\bibnamefont {Bergemann}}, \bibinfo {author}
  {\bibfnamefont {B.~J.}\ \bibnamefont {Daurer}}, \bibinfo {author}
  {\bibfnamefont {T.}~\bibnamefont {Ekeberg}}, \bibinfo {author} {\bibfnamefont
  {A.~D.}\ \bibnamefont {Estillore}}, \bibinfo {author} {\bibfnamefont
  {H.}~\bibnamefont {Fangohr}},  \emph {et~al.},\ }\href@noop {} {\bibfield
  {journal} {\bibinfo  {journal} {IUCrJ}\ }\textbf {\bibinfo {volume} {9}}
  (\bibinfo {year} {2022})}\BibitemShut {NoStop}%
\bibitem [{\citenamefont {Zhong}\ \emph {et~al.}(2021)\citenamefont {Zhong},
  \citenamefont {Bepler}, \citenamefont {Berger},\ and\ \citenamefont
  {Davis}}]{Zhong:2021}%
  \BibitemOpen
  \bibfield  {author} {\bibinfo {author} {\bibfnamefont {E.~D.}\ \bibnamefont
  {Zhong}}, \bibinfo {author} {\bibfnamefont {T.}~\bibnamefont {Bepler}},
  \bibinfo {author} {\bibfnamefont {B.}~\bibnamefont {Berger}}, \ and\ \bibinfo
  {author} {\bibfnamefont {J.~H.}\ \bibnamefont {Davis}},\ }\href@noop {}
  {\bibfield  {journal} {\bibinfo  {journal} {Nature methods}\ }\textbf
  {\bibinfo {volume} {18}},\ \bibinfo {pages} {176} (\bibinfo {year}
  {2021})}\BibitemShut {NoStop}%
\end{thebibliography}%
\end{document}